\documentstyle[aps,preprint,epsf]{revtex}

\title{Ising fluids in an external magnetic field: \\
       an integral equation approach}

\author{I. P. Omelyan,$^{1,2}$ I. M. Mryglod,$^{1,2}$
        R. Folk,$^2$ and W. Fenz$^2$}

\address{$^1$Institute for Condensed Matter Physics,
         1 Svientsitskii Street, UA-79011 Lviv, Ukraine}
\address{$^2$Institute for Theoretical Physics, Linz University,
         A-4040 Linz, Austria}

\date{\today}

\begin{document}

\maketitle

\begin{abstract}

The phase behavior of Ising spin fluids is studied in the presence of an
external magnetic field with the integral equation method. The calculations
are performed on the basis of a soft mean spherical approximation using an
efficient algorithm for solving the coupled set of the Ornstein-Zernike
equations, the closure relations, and the external field constraint. The
phase diagrams are obtained in the whole thermodynamic space including the
magnetic field $H$ for a wide class of Ising fluid models with various ratios
$R$ for the strengths of magnetic to nonmagnetic Yukawa-like interactions.
The influence of varying the inverse screening lengths $z_1$ and $z_2$,
corresponding to the magnetic and nonmagnetic Yukawa parts of the potential,
is investigated too. It is shown that changes in $R$  as well as in $z_1$ and
$z_2$ can lead to different topologies of the phase diagrams. In particular,
depending on the value of $R$, the critical temperature of the liquid-gas
transition either decreases monotonically, behaves nonmonotonically, or
increases monotonically  with increasing $H$. The para-ferro magnetic
transition is also affected by changes in $R$ and the screening lengths. At
$H=0$, the Ising fluid maps onto a simple model of a symmetric nonmagnetic
binary mixture. For $H \to \infty$, it reduces to a pure nonmagnetic fluid.
The results are compared with available simulations and the predictions of
other theoretical methods. It is demonstrated, that the mean spherical
approximation appears to be more accurate compared with  mean field theory,
especially for systems with short ranged attraction potentials (when $z_1$
and $z_2$ are large). In the Kac limit $z_1,z_2 \to +0$, both approaches tend
to nearly the same results.

\vspace{6pt}

\noindent
PACS number(s): 64.70.Fx, 05.70.Fh, 64.60.-i, 75.50.Mm

\end{abstract}

\vspace{24pt}

\section{Introduction}

During the past decades, much attention has been paid to the phase behavior
of ferromagnetic fluid models with coupled spin and spatial interactions
[1--20]. The investigations have been carried out using the mean field (MF)
theory \cite{Frankel,HemImb77,Tavares,Schinagl,Schinfol,Fenz,Fefomel}, the
method of integral equations \cite{Lom,Sokolovska,Lados,Lado,Sokolovskii,%
Lomba}, as well as Monte Carlo (MC) simulation techniques \cite{Fefomel,%
Lom,Lado,Nielaba,NijWe,Nijmeijer,Weis,Parola,Ferreira,Mryglod}. They dealt
mainly with simplified models belonging to a so-called ideal class of spin
fluids, where the attractive part of nonmagnetic interactions is absent
($R=\infty$). Moreover, these models were considered, as a rule, in the
absence of an external magnetic field ($H=0$). At finite values of $R$, it
has been established \cite{Frankel,HemImb77,Schinagl,Schinfol} that,
depending on the relative strength of magnetic interactions, the gas, liquid,
paramagnetic, and ferromagnetic states in the system may form  phase diagrams
of different topologies. For instance, an order-disorder liquid-liquid phase
transition may appear additionally to the gas-liquid one.

A complete picture on the phase diagram topology can only be obtained if the
field is included ($H \ne 0$) in the consideration. Then the phase diagram
shows the two other critical lines (so called wings) meeting the magnetic
transition line in the $H=0$ plane at the tricritical liquid-liquid
transition \cite{Schinagl,Schinfol}. The gas-liquid critical point extends to
a critical line in the magnetic field. Whether the gas-liquid or
liquid-liquid critical line ends in a critical end point and the
corresponding other critical line tends to infinite magnetic field depends on
the model parameters mentioned. This enlarges the number of different
``phases'' in the global phase diagram (the regions where the different
topologies exist in the space of the microscopic model parameters, like
interaction strength or range of the potentials).

The global phase diagrams in one-dimensional $R$-domain were obtained for
Ising and Heisenberg spin fluids within MF theory \cite{Schinagl,Schinfol}.
To our knowledge, no such diagrams have been investigated up to now using the
integral equation method. As is well recognized, the latter method leads to
more accurate predictions. It takes into account pair correlations between
particles in spin space, which are completely ignored by the MF approach.
Previous integral equation studies on magnetic fluids have been restricted
exclusively to ideal systems with Heisenberg spin interactions in the absence
of an external field \cite{Lom,Sokolovska} or only included a few nonzero
field values \cite{Lados,Lado,Sokolovskii,Lomba}. No integral equation
calculations have been performed for nonideal spin fluids at $H \ne 0$, even
within the well-known Ising model. Note that here we are dealing with genuine
fluid models, meaning that the spatial positions of spins are distributed
continuously, contrary to simplified lattice gas schemes \cite{Kawasaki,%
Sokolovski,Romano}, where the spins are positioned on fixed lattice sites.

Due to the discrete character of spin reorientations in the Ising
fluid, it can be mapped onto a binary nonmagnetic mixture with
symmetric interparticle interactions. In this context, it should
be pointed out that in recent years a lot of papers have been
devoted to study phase properties in symmetric mixtures by means
of MC simulations, the MF theory, as well as the Ornstein-Zernike
(OZ) integral equation method
\cite{Wilding,Kahlang,Paschingerlev,Antonevych,Kahl,Paschinger,Pini,Wildinga}.
Various closure relations, including the standard mean spherical
approximation (MSA) \cite{Hansen} and a self-consistent OZ
approach (SCOZA) \cite{Paschinger,Pinis}, have been exploited
within the latter method. Note that these studies considered in
fact only the case when the chemical potentials of different
species are fixed to be equal. In the language of Ising fluids
this corresponds to the absence of an external magnetic field.

It has been realized that despite its relative simplicity, the MSA
is able to give reliable results for the coexistence phase
boundaries including the location of critical points. The accuracy
of the MSA scheme gets worse only when calculating critical
exponents. On the other hand, the more cumbersome SCOZA technique
can provide us with highly precise results for the phase
boundaries and it remains accurate even near criticality
\cite{Pinis}. The other concept is the hierarchical reference
theory (HRT) \cite{Pini,Parolare,Tau,Brognara}, that combines
features of the renormalization group theory (RGT) and theoretical
liquid-state approaches and allows to reproduce some critical
exponents more precisely with respect to the MSA. For instance,
the critical exponent $\beta$ (which gives the curvature of the
coexistence curve near the critical point) takes the values 7/20
and 0.345 within the SCOZA and HRT, respectively. They are close
to the experimental and RGT prediction $\beta=0.327$, contrary to
the MF and MSA value $\beta=1/2$ (see section III.B).

However, the SCOZA has so far been implemented only for a restricted class of
hard-sphere-Yukawa potentials. For these potentials, some solutions within
the MSA ansatz can be derived \cite{Blum,Arrieta,Tang} in a semianalytical
form as a set of nonlinear algebraic equations (which should further be
solved numerically). The SCOZA employs such solutions at an intermediate
stage of the calculations. In the presence of potentials of any other
structure, for instance, of Lennard-Jones (LJ) type or potentials with a
soft-core repulsion part, the mathematical structure is less amenable. In
addition, the SCOZA enforces the consistency between different thermodynamic
routes in a somewhat phenomenological manner by introducing an artificial
``temperature'' depending on density and concentration. The high level of
sophistication of the SCOZA and HRT concepts leads, in turn, to substantial
computational problems, when applying them to more realistic interaction
potentials. Therefore, for our problem we stick to a variant of the MSA,
being well aware of the properties of this approximation near the critical
point.

In the present paper, the global phase diagram of the Ising fluid are
investigated on the basis of the OZ integral equation method with a soft MSA
closure. This allows us to obtain the complete thermal phase diagrams
covering the whole range of the relative strength of magnetic interactions
and other parameters of the interaction potentials. The dependencies of the
critical temperatures and densities on the external field are analyzed in
details as well.

\section{Background}

\subsection{The Ising model}

The full potential energy of the Ising fluid can be written in
its most general form as
\begin{equation} \label{eq1}
U=\frac{1}{2} \sum_{i \ne j}^{N} \Big[ \varphi(r_{ij}) -
I(r_{ij}) - J(r_{ij}) \, s_i s_j \Big] - H \sum_{i=1}^{N} s_i \, ,
\end{equation}
where ${\bf r}_i$ is the (three-dimensional) spatial coordinate of the
$i$-th particle carrying spin $s_i=\pm 1$ with $i=1,\ldots, N$ and $N$
being the total number of particles, $r_{ij}=|{\bf r}_i-{\bf r}_j|$
denotes the interspin separation, and $H$ relates to the homogeneous
external magnetic field. The exchange integral ($J>0$) of ferromagnetic
interactions and the attraction part ($I>0$) of nonmagnetic ones can be
chosen in the form of Yukawa functions,
\begin{eqnarray} \label{eq2}
J(r) = \frac{2(z_1 \sigma)^2}{z_1 \sigma+1} \frac{\epsilon_J \sigma}{r}
\exp[-z_1(r-\sigma)] \, , \nonumber \\ [-5pt] \\ [-5pt]
I(r) = \frac{2(z_2 \sigma)^2}{z_2 \sigma+1} \frac{\epsilon_I \sigma}{r}
\exp[-z_2(r-\sigma)] \, , \nonumber
\end{eqnarray}
where $z_1$ and $z_2$ are the inverse screening lengths of the potentials,
$\epsilon_J$ and $\epsilon_I$ denote the interaction intensities, and
$\sigma$ relates to the particle size. The repulsion $\varphi$ between
particles can be modeled by a (more realistic) LJ-like soft-core (SC)
potential \cite{Fefomel},
\begin{equation} \label{eq3}
\varphi(r) = \left\{
\begin{array}{cc}
\displaystyle
4 \varepsilon \bigg[ \Big( \frac{\sigma}{r}
\Big)^{12} - \Big( \frac{\sigma}{r} \Big)^6 \bigg] +
\varepsilon \, , & \ \ r < \sqrt[6]{2} \sigma \, , \\ [12pt]
0 \, , & \ \ r \ge \sqrt[6]{2} \sigma \, ,
\end{array}
\right.
\end{equation}
rather than by the hard-sphere (HS) function
\begin{equation} \label{eq4}
\varphi_{\rm HS}(r) = \left\{
\begin{array}{rl}
\displaystyle
\infty \, , & \ \ \ r < \sigma \, , \\ [12pt]
0 \, \, , & \ \ \ r \ge \sigma \, .
\end{array}
\right.
\end{equation}

The multipliers $2 (z_{1,2} \sigma)^2/(z_{1,2} \sigma+1)$, entering in
Eq.~(\ref{eq2}), have been used for the sake of convenience of comparison of
our results with previous predictions (see section III). Then, for instance,
the integrals $\int_\sigma^\infty I(r) {\rm d} {\bf r} = 8 \pi \epsilon_I
\sigma^3$ and $\int_\sigma^\infty J(r) {\rm d} {\bf r} = 8 \pi \epsilon_J
\sigma^3$  are independent of $z_1$ and $z_2$, respectively. Within the
standard hard-sphere MF theory (HSMF), such integrals describe the
contribution of the interactions to the free energy \cite{Tavares}. Thus, we
can say in advance that the HSMF results will not dependent on $z_{1,2}$. At
$z_{1,2} \sigma=1$ (the case which is usually considered in theory and
simulation), the multipliers go to unity, and we come to the usual form for
$I(r)$ and $J(r)$. Within the soft-core MF theory (SCMF), introduced recently
in Ref.~\cite{Fefomel}, a slight $z_{1,2}$-dependency should appear. Then the
integrals transform to $\int_0^\infty \exp(-\beta \varphi_{\rm SC})
\{I,J\}(r) {\rm d} {\bf r} = 8 \gamma(T,z_{1,2}) \pi \epsilon_{I,J}
\sigma^3$, where $\beta^{-1}=k_{\rm B} T$ denotes the temperature with
$k_{\rm B}$ being Boltzmann's constant, and $\gamma(T,z_{1,2})$ is the
function which takes into account the softness of repulsion potentials (see
Eqs.~(8)-(10) of Ref.~\cite{Fefomel}). For the integral equation approach we
expect a more pronounced dependence of the results on $z_{1,2}$.

\subsection{Integral equation approach}

\subsubsection{Mapping to a symmetric binary mixture}

Since the spins $s_i$ in an Ising fluid take only two values,
$\pm1$, we can map the system with $N$ particles carrying spin
$1$ or $-1$ onto a binary mixture with $N_a$ and $N_b$ particles
of type $a$ and $b$, respectively, where $N_a+N_b=N$.
Then Eq.~(1) transforms to the equivalent form
\begin{equation} \label{eq5}
U\!=\frac{1}{2} \sum_{i \ne j}^{N_{a}} \phi_{aa}(r_{ij}) \!+\!
\frac{1}{2} \sum_{i \ne j}^{N_{b}} \phi_{bb}(r_{ij}) \!+\!\!
\sum_{i,j=1}^{N_{a},N_{b}} \phi_{ab}(r_{ij}) - H M
\end{equation}
where $M=\sum_{i=1}^{N} s_i=N_a-N_b$ relates to the total magnetization
of the system and
\begin{eqnarray} \label{eq6}
\phi_{aa}(r)&=&\phi_{bb}(r)=\varphi(r) - [I(r) + J(r)] \, ,
\nonumber \\ [-12pt] \\ [-12pt]
\phi_{ab}(r)&=&\phi_{ba}(r)=\varphi(r) - [I(r) - J(r)]
\nonumber
\end{eqnarray}
describe the interactions between like and unlike particles in the mixture.

In a further step, we have to rewrite the energy, the magnetic field, and the
magnetization per particle $m=M/N$ in terms of appropriate variables suitable
for the mixture. Apart from the total number density $\rho=N/V$, where $V$
denotes the volume, such variables include the particle concentration $x$ and
the chemical potentials. One has the concentration relations
\begin{equation} \label{concmag}
x=\frac{N_a}{N} \, , \qquad 1-x=\frac{N_b}{N} \, , \qquad m=2x-1 \, .
\end{equation}
In addition, we emphasize that due to a finite value of the external field
term on the right-hand side of Eq.~(\ref{eq5}), we have to deal with a
mixture prepared in an unusual way. Indeed, when transferring one particle
from species $a$ to species $b$ -- in the Ising liquid this amounts to
flipping a spin from up to down and thus $\Delta M=-2$ -- without changing
the spatial coordinates, the total change in energy is equal to $\Delta U=2H$
(the other terms do not contribute to $\Delta U$ due to the symmetry
$\phi_{aa}=\phi_{bb}$ of the particle interactions). On the other hand, the
change in energy of the mixture is given by
\begin{equation}
\Delta U=\Delta N_a\mu_a+\Delta N_b\mu_b=\mu_b(\rho,x,T)-\mu_a(\rho,x,T) \, ,
\end{equation}
where $\mu_a$ and $\mu_b$ are the chemical potentials of species $a$ and $b$,
respectively. This leads to the identification
\begin{equation} \label{eq14}
\Delta\mu\equiv\mu_b(\rho,x,T)-\mu_a(\rho,x,T)=2H \, .
\end{equation}
Condition (\ref{eq14}) can be considered as an additional constraint imposed
on the concentration at given values of $\rho$, $T$, and $H$, namely,
$x=x(\rho,T,H)$. The reason for this procedure is the following. In the case
of a magnetic fluid, the external field (not the magnetization) is accessible
to experiment, whereas for a mixture it is the concentration (note the
chemical potentials). In order to study the Ising fluid in the notation of
binary mixture one, therefore, has to fix the difference of the chemical
potentials.

\subsubsection{Formulation of the integral equations}

For mixtures, the OZ integral equations have \cite{Hansen} the form
\begin{equation} \label{eq7}
h_{\alpha \beta}(r) = c_{\alpha \beta}(r) + \sum_{\gamma=a,b} \rho_\gamma
\int c_{\alpha \gamma}(|{\bf r}-{\bf r'}|) \, h_{\gamma \beta}(r') \,
{\rm d} {\bf r'} \, ,
\end{equation}
where the total $h_{\alpha \beta}$ and direct $c_{\alpha \beta}$ correlation
functions for a pair of particles of species $\alpha$ and $\beta$ will depend
only on their separation distance, $\rho_\gamma=N_\gamma/V$ is the particle
number density of the $\gamma$-th species, and the indices $\alpha$, $\beta$
and $\gamma$ accept two values, $a$ (spin up) and $b$ (spin down).

Eq.~(\ref{eq7}) must be complemented by an (approximate) closure relation to
be solved with respect to $h_{\alpha \beta}$ and $c_{\alpha \beta}$. The
standard MSA scheme \cite{Hansen}, proposed originally \cite{Lebowitz} for
systems with HS repulsion (\ref{eq4}), should be replaced in our case by the
soft MSA ansatz (SMSA) \cite{Madden,Choudhury}, appropriate for a SC
potential (\ref{eq3}). It reads
\begin{equation} \label{eq8}
g_{\alpha \beta}(r) = \exp[-\beta \phi_{\alpha \beta}(r) + h_{\alpha
\beta}(r)-c_{\alpha \beta}(r)+B_{\alpha \beta}(r)] \, ,
\end{equation}
where $g_{\alpha \beta}(r)=h_{\alpha \beta}(r)+1$ denotes the radial
distribution function, and
\begin{equation} \label{eq9}
B_{\alpha \beta}(r)=\ln[1+s_{\alpha \beta}(r)]-s_{\alpha \beta}(r)
\end{equation}
is the bridge function with
\begin{equation} \label{eq10}
s_{\alpha \beta}(r)=h_{\alpha \beta}(r)-c_{\alpha \beta}(r)-
\beta \phi_{\alpha \beta}^{\rm l}(r)
\end{equation}
(no confusion may arise between the index and the Boltzmann factor $\beta$).
Formally setting $B=0$ in Eq.~(\ref{eq8}) leads to the hypernetted-chain
(HNC) approximation \cite{Hansen}.

The SMSA additionally requires the separation of the total potential
$\phi_{\alpha \beta}$ in its short- and long-ranged parts $\phi_{\alpha
\beta}^{\rm s}$ and $\phi_{\alpha \beta}^{\rm l}$ with $\phi_{\alpha \beta}=
\phi_{\alpha \beta}^{\rm s}+\phi_{\alpha \beta}^{\rm l}$. There is no general
procedure to perform such a separation uniquely for arbitrary potentials.
Since the SMSA itself is not exact, the Yukawa potentials in the region of
core repulsion $r \lesssim \sigma$ allow splitting to some extent in various
ways, leading to various versions of the SMSA. Usually the splitting is
carried out by introducing a switch function. One choice among others is to
extract the long-ranged part using the Boltzmann-like switching exponent
built on the soft-core potential, i.e.,
\begin{equation}
\phi_{\alpha \beta}^{\rm l}(r) = - [I(r) \pm J(r)] \exp[-\beta \varphi(r)]
\, .
\end{equation}
Such an extraction is quite natural, because for $r > \sigma$ the function
$\phi_{\alpha \beta}^{\rm l}(r)$ rapidly tends with increasing $r$ to the
Yukawa potential $-[I(r) \pm J(r)]$ (we note that $\exp[-\beta \varphi(r)]=1$
for $r \ge \sqrt[6]{2} \sigma$, whereas $\lim_{r \to 0} \exp[-\beta \varphi
(r)]=0$). By the replacement $\varphi(r) \to \varphi_{\rm HS}(r)$, we come to
the standard HS MSA with $h_{\alpha \beta}(r) = 0$ for $r < \sigma$ and
$c_{\alpha \beta}(r) = \beta [I(r) \pm J(r)]$ for $r \ge \sigma$.

Another trick lies in a modification of the bridge function to the form of
Eq.~(\ref{eq9}) if $s_{\alpha\beta}(r)>0$ and to $B_{\alpha\beta}(r)=0$ when
$s_{\alpha\beta}(r) \le 0$, that combines the SMSA with the HNC
approximation. This is in the spirit of the KH closure proposed by Kovalenko
and Hirata \cite{Kovalenkos,KovHirata}. Note that the pure SMSA sometimes
leads to unphysical $r$-domains with negative values of $g(r)$. The modified
SMSA preserves by construction the positiveness of $g(r)$ everywhere in
density and temperature space.

\subsubsection{Calculation of thermodynamic quantities}

Eqs.~(\ref{eq7}) and (\ref{eq8}) constitute a system of six nonlinear
integro-algebraic equations for the same number of unknowns $\{h,c\}_{aa}$,
$\{h,c\}_{bb}$, and $\{h,c\}_{ab}=\{h,c\}_{ba}$. Once the solutions are
found, the thermodynamic quantities are calculated in a straightforward way.
In particular, the pressure can be calculated from the virial equation of
state
\begin{equation} \label{eq13}
P(\rho,x,T) = \rho k_{\rm B} T - \frac{2 \pi}{3} \sum_{\alpha,\beta}^{a,
b} \rho_\alpha \rho_\beta \int_0^\infty \frac{{\rm d} \phi_{\alpha
\beta}}{{\rm d} r} g_{\alpha \beta}(r) r^3 {\rm d} r \, ,
\end{equation}
where $\rho=\rho_a+\rho_b$. Although the energy and compressibility routes
can also be used, we will prefer the virial route (\ref{eq13}) because it is
most easily implemented.

The chemical potentials can be written in the form
\begin{equation} \label{eq15}
\mu_\alpha = \mu_\alpha^\ast + k_{\rm B} T ( \ln \varrho_\alpha +
3 \ln \Lambda_\alpha)
\end{equation}
where $\alpha=a,b$ and $\Lambda_\alpha$ being the de Broglie thermal
wavelength (which is independent of $\rho$ and $x$). Explicit
expressions for the excess part of $\mu_\alpha$ can be derived
using the (exact) Kirkwood formula \cite{Kirkwood}
\begin{equation}
\mu_\alpha^\ast = \sum_{\beta=a,b} \rho_\beta \int_0^1 {\rm d} \lambda
\int_0^\infty g_{\alpha \beta}(r,\lambda) \frac{\partial \varphi_{\alpha \beta}(r,
\lambda)}{\partial \lambda} 4 \pi r^2 {\rm d} r \, .
\end{equation}
Here, the integration over $\lambda$ corresponds to the computation of the
work of transferring a separate particle from a vacuum ($\lambda=0$ with
$\varphi_{\alpha \beta}(r,\lambda)=0$) to the system ($\lambda=1$ and
$\varphi_{\alpha \beta}(r,\lambda)= \varphi_{\alpha \beta}(r)$). Performing
the $\lambda$-integration in a manner similar to that proposed in
Refs.~\cite{Kovalenkos,KovHirata}, one obtains, taking into account
Eqs.~(\ref{eq7}) and (\ref{eq8}) that
\begin{eqnarray} \label{eq16}
\mu_\alpha^\ast = && k_{\rm B} T \sum_{\beta=a,b} \rho_\beta \int_0^\infty
\bigg[ \frac12 h_{\alpha \beta}^2(r) - \frac12 h_{\alpha \beta}(r) c_{\alpha
\beta}(r) - c_{\alpha \beta}(r) \nonumber \\ [-6pt]
\\ [-6pt] && + B_{\alpha \beta}(r) g_{\alpha \beta}(r) - \frac{h_{\alpha
\beta}(r)}{s_{\alpha \beta}(r)} \int_0^{s_{\alpha \beta}(r)} B(s') {\rm d} s'
\bigg] 4 \pi r^2 {\rm d} r \, , \nonumber
\end{eqnarray}
with $\int_0^s \! B(s') {\rm d} s'$ being equal to $(1+s) \ln(1+s) -
s(s+2)/2$ at $s > 0$ or 0 for $s \le 0$.

It is worth mentioning that since the correlation functions are
calculated approximately, the above expressions (15) and (18) for
the pressure and chemical potentials will not be thermodynamically
self-consistent. In particular, the virial, energy and
compressibility routes will lead to different results. Although
the differences are, as a rule, relatively small, they may distort
the phase coexistence properties near a critical point. A
derivation of the self-consistent expressions in the case of the
nonideal Ising model at the presence of the external field is a
non-trivial problem that needs a separate serious investigation.
It could be solved, for instance, within the SCOZA by extending
its present implementation from hard-sphere to soft-core repulsion
potentials. In this paper for the sake of simplicity we will use
the virial pressure complemented by an appropriately chosen
Maxwell construction (see subsection II. C).

\subsection{Phase separations}

For a general binary mixture, the densities $\rho^{\rm I,II}$ and
concentrations $x^{\rm I,II}$ of coexisting phases I and II are determined at
a given temperature $T$ from the well-known mechanical and chemical
equilibrium conditions
\begin{eqnarray} \label{eq19}
P\big(\rho^{\rm I},x^{\rm I},T\big)&=&
P\big(\rho^{\rm II},x^{\rm II},T\big) \, , \nonumber \\
\mu_a\big(\rho^{\rm I},x^{\rm I},T\big)&=
&\mu_a\big(\rho^{\rm II},x^{\rm II},T\big) \equiv \mu_a^{\rm I,II} \, ,\\
\mu_b\big(\rho^{\rm I},x^{\rm I},T\big)&=&
\mu_b\big(\rho^{\rm II},x^{\rm II},T\big) \equiv \mu_b^{\rm I,II} \, .
\nonumber
\end{eqnarray}
In our case, they should be complemented by the condition
\begin{equation}
\mu^{\rm I,II}_b-\mu^{\rm I,II}_a=2H
\end{equation}
following from the external field constraint (Eq.~(\ref{eq14})) for each
phase. It is convenient to replace $\mu_a$ and $\mu_b$ by the sum
\begin{equation}
\mu\equiv\frac{\mu_a+\mu_b}{2}
\end{equation}
and the difference $\Delta \mu$ defined in (\ref{eq14}). Then two of the
three chemical potential conditions in Eqs.~(\ref{eq19}) and (20) can be
rewritten in the equivalent form
\begin{equation}
\Delta \mu\big(\rho^{\rm I},x^{\rm I},T\big)=
\Delta \mu\big(\rho^{\rm II},x^{\rm II},T\big)=2H \, .
\end{equation}

These conditions will be satisfied automatically, provided the
integro-algebraic equations (\ref{eq7}) and (\ref{eq8}) are solved in
conjunction with Eq.~(\ref{eq14}). Then one finds a consistent set of
correlation functions together with the solution
\begin{equation} \label{eq21}
x=x(\rho,T,H)
\end{equation}
for the concentration in the mixture. For the Ising fluid, the magnetization
$m(\rho,T,H)$ can easily be reproduced from $x$, whenever it is necessary,
using relation (\ref{concmag}). Solution (\ref{eq21}) can now be inserted
into the remaining conditions yielding
\begin{eqnarray} \label{eq22}
P\big(\rho^{\rm I},T,H\big)&=&P\big(\rho^{\rm II},T,H\big) \, ,
\nonumber \\ [-12pt] \\ [-12pt]
\mu\big(\rho^{\rm I},T,H\big)&=&\mu\big(\rho^{\rm II},T,H\big) \, , \nonumber
\end{eqnarray}
where the mapping from $(\rho,x,T)$-space to the new set
$\big(\rho,x(\rho,T,H),T\big) \equiv (\rho,T,H)$ has been performed.

Relations (\ref{eq22}) look now like the coexistence conditions for a
one-component fluid. Indeed, the Gibbs free energy of the mixture $G = \mu_a
N_a + \mu_b N_b$ can be rewritten in terms of $N=N_a+N_b$ and $M=N_a-N_b$ as
$ G = \frac{\mu_a+\mu_b}{2} N - \frac{\mu_b-\mu_a}{2} M \equiv \mu N - H M $,
so that the introduced quantity $\mu$ (Eq.~(21)) has the meaning of the
chemical potential of the Ising fluid. It can be calculated using expressions
(\ref{eq15}) and (\ref{eq16}) [which were already used when solving the
external field constraint (see Eqs.~(9) and (23)]. Alternatively, a
Maxwell-construction scheme has been utilized. In order to demonstrate that
this scheme can be applied in its standard form to the magnetic fluid, let us
consider the change in the Gibbs free energy. According to the thermodynamic
relations, we have ${\rm d} G = - S {\rm d} T + V {\rm d} P + \mu_a {\rm d}
N_a + \mu_b {\rm d} N_b \equiv - S {\rm d} T + V {\rm d} P + \mu {\rm d} N -
H {\rm d} M$, where $S$ denotes the entropy. During the isothermal (${\rm d}
T = 0$) process ${\rm I} \to {\rm II}$ we obtain, integrating by parts, that
for the system with a fixed number of particles (${\rm d} N=0$) in the
presence of a constant external field $H$, the free energy increment is equal
to $\Delta G = PV|_{\rm I}^{\rm II} - \int_{\rm I}^{\rm II} P {\rm d} V - H
(M_{\rm II}-M_{\rm I})$. On the other hand, from the definition of $G$ it
follows that $\Delta G = N \mu|_{\rm I}^{\rm II} - H (M_{\rm II}-M_{\rm I})$.
Since the increment caused by the change $H (M_{\rm II}-M_{\rm I})$ in the
magnetic energy is the same for both routes, we come to the Maxwell
construction $ Q_{\rm I,II}=(1/\rho^{\rm II} - 1/\rho^{\rm I})\,{\cal P} +
\int_{\rho^{\rm I}}^{\rho^{\rm II}} P(\rho,T,H) {\rm d} \rho/\rho^2=0$, where
${\cal P}$ denotes the coexistence pressure. The construction guarantees that
the chemical potential will be the same ($\mu|_{\rm I}^{\rm II}=0$) in both
phases I and II, so that the second line of Eq.~(24) transforms to
$Q\big(\rho^{\rm I},\rho^{\rm II},T,H\big)=0$.

In such a way, the gas-liquid and liquid-liquid phase transitions of the
first order can be determined. The second-order para-ferro magnetic
transition at $H=0$ can be found as a boundary (Curie) curve
$T_\lambda(\rho)$. Below this curve, i.e., for $T < T_\lambda(\rho)$,
Eq.~(\ref{eq14}), being solved at $H=0$ with respect to the concentration $x$
(see Eq.~(\ref{eq21})), should have a nontrivial solution $x=x(\rho,T,0) \ne
1/2$ (i.e., $m \ne 0$). For $T > T_\lambda(\rho)$, only the trivial one
should satisfy Eq.~(\ref{eq14}) at $H=0$. Note that the trivial solution
$x=1/2$ (or $m=0$) appears as a result of the symmetry of the interparticle
potentials (Eq.~(\ref{eq6})). For the same reason, the coexistence phases
will have an identical density at concentrations $x$ and $1-x$, or at
magnetizations $m$ and $-m$. From the structure of Eqs.~(\ref{eq14}),
(\ref{eq13}) and (\ref{eq16}) it also follows that, if $x$ (or $m$) is a
solution to equation (\ref{eq14}) at some value of $H$, then $1-x$ (or $-m$)
will automatically satisfy this equation at the field $-H$. Therefore, the
phase diagrams will be symmetric with respect to the magnetic field $H$.

\section{Numerical calculations}

\subsection{Computational algorithm}

The set of OZ integral equations (\ref{eq7}) was first reduced to a system of
linear algebraic ones, $h_{\alpha \beta}(k) = c_{\alpha \beta}(k) +
\sum_{\gamma=a,b} \rho_\gamma c_{\alpha \gamma}(k) \, h_{\gamma \beta}(k)$,
by applying the 3-dimensional Fourier transform $A(k)=\int_V A(r) \exp({\rm
i} {\bf k \cdot \bf r}) {\rm d} {\bf r}=\int_0^\infty 4 \pi r^2 A(r)
\sin(kr)/(kr) {\rm d} r$, where $A$ is any function of $r$. It can be
presented in the compact ($2\times2$) matrix form ${\bf h}(k) = {\bf c}(k) +
{\bf c}(k) {\mbox{\boldmath $\rho$}} {\bf h}(k)$ with ${\mbox{\boldmath
$\rho$}}$ being the diagonal density matrix having nonzero elements
[${\mbox{\boldmath $\rho$}}]_{11}=\rho_a \equiv x\rho$ and [${\mbox{\boldmath
$\rho$}}]_{22}=\rho_b \equiv (1-x)\rho$.

Because of the nonlinearities in the SMSA closure (Eqs.~(11)--(14)) and the
external field constraint (FC) [see Eqs.~(\ref{eq14}) and (\ref{eq16})], the
coupled set of OZ/SMSA/FC equations have to be solved iteratively. The
iterations have been carried out by adapting the method of modified direct
inversion in the iterative subspace (MDIIS) \cite{Mdiis}. At given values of
$\rho$, $T$, and $H$ the iteration starts from initial guesses for $c_{\alpha
\beta}(r)$ and $x$, and the Fourier transformed functions $c_{\alpha
\beta}(k)$ are calculated. Then the total correlation functions in $k$-space
are obtained analytically, ${\bf h}(k)=[{\bf I}-{\bf c}(k) {\mbox{\boldmath
$\rho$}}]^{-1} {\bf c}(k)$, where ${\bf I}$ denotes the unit matrix. Applying
the backward Fourier transform to ${\bf h}(k)$ yields $ {\bf h}(r)=1/(2\pi)^3
\int_0^\infty 4 \pi k^2 {\bf h}(k) \sin(kr)/(kr) {\rm d} k $. With the
current values of ${\bf c}(r)$, ${\bf h}(r)$, and $x$, the residuals to the
SMSA closure (\ref{eq8}) and field constraint (\ref{eq14}) are evaluated.
Using them, new values of ${\bf c}(r)$ and $x$ are updated according to the
MDIIS corrections, and the iteration procedure is repeated until the
solutions lead to residuals with a relative root mean square magnitude of
$10^{-6}$. The coexistence phase densities were then found by applying the
Maxwell construction.

The ratio $R$ of the integrated strengths of magnetic to nonmagnetic
interactions can be calculated as
\begin{equation} \label{eq27}
R=\frac{\int_0^\infty 4 \pi g(r) J(r) r^2 {\rm d} r}
{\int_0^\infty 4 \pi g(r) I(r) r^2 {\rm d} r} \, ,
\end{equation}
where $g(r) = x^2 g_{aa}(r) + 2 x (1-x) g_{ab}(r) + (1-x)^2 g_{bb}(r)$ is the
(total) radial distribution function of the Ising fluid. From the form of
Yukawa potentials $I(r)$ and $J(r)$ (Eq.~(\ref{eq2})) it follows that the
relation (\ref{eq27}) transforms to
\begin{equation} \label{ratio}
R=\frac{\epsilon_J}{\epsilon_I}
\end{equation}
for $z_1=z_2$, We have preferred, the latter definition even when $z_1 \ne
z_2$, because the former is sensitive to the approximation made for $g(r)$.

The strength $\varepsilon$ appearing in the SC potential (Eq.~(\ref{eq3}))
was set to $\epsilon_J$. This corresponds to a moderate softness of $\varphi$
with respect to the total potential (see Eq.~(\ref{eq1})). In the
presentation of our results we use the dimensionless density $\rho^\ast=\rho
\sigma^3$, temperature $T^\ast=k_{\rm B} T/\epsilon_J$, external field
$H^\ast=H/ \epsilon_J$, and inverse screening lengths
$z^\ast_{1,2}=z_{1,2}\sigma$.

\subsection{Results for the ideal system $(R=\infty)$}

Examples of the phase diagrams obtained within the OZ/SMSA/FC integral
equation approach for the soft-core ideal Ising fluid with $z_1^\ast=1$ at
various values, $H^\ast= 0$, 0.1, 0.5, 1, 5, and $\infty$, of the external
field are shown in sets (a), (b), (c), (d), (e), and (f) of Fig.~1,
respectively, in the ($T^\ast,\rho^\ast$) plane. For the purpose of
comparison, the results of the SCMF theory and available MC simulation data
\cite{Fefomel} are also included in this figure.

As can be seen clearly, the OZ/SMSA/FC approach leads to much more accurate
predictions of the liquid-gas coexistence densities with respect to the usual
version of the SCMF theory. Even the adjustable version, when a
semiphenomenological parameter is introduced within the SCMF and fitted to MC
data at $H \to \infty$ (see Ref.~\cite{Fefomel}), provides us with worse
results. At the same time, the deviations between the OZ/SMSA/FC predictions
and MC data are relatively small, especially for regions which are well below
the critical point. On the other hand, the precision decreases when
approaching the criticality point, where the uncertainties in critical
temperature and density estimations can reach about 10--15\%. Moreover, the
computations have shown that the OZ/SMSA/FC approach reproduces, like the MF
theory, the classical values of critical exponents. For instance, the density
difference in liquid and gas phases, $\rho_{\rm L}^\ast - \rho_{\rm G}^\ast$,
appears to be proportional near a critical point to $(T_{\rm c}^\ast-
T^\ast)^{\beta}$ independently of $H^\ast$ with the critical exponent
$\beta=1/2$ (instead of the values $\beta \approx 1/3$ and $7/20$ obtained
within the RGT and SCOZA, respectively).

The influence of varying the screening length $z_1$ of magnetic interactions
on the OZ/SMSA/FC phase diagram is illustrated in Fig.~2. In this respect, it
should be emphasized that within the SCMF scheme, the results will depend on
$z_1$ very weakly (see comments at the end of subsection II.A). Using the
more precise integral equation approach, we can observe an obvious
$z$-dependence of the binodal for all values of the external field. In
particular, the dimensionless critical temperature $T_{\rm c}^\ast$ and
density $\rho_{\rm c}^\ast$ increase considerably with rising $z_1$ at each
fixed value of $H$. However, the topology of the phase diagram remains the
same and is not affected by varying the screening length. Namely, as in the
case of the SCMF theory \cite{Fefomel}, the OZ/SMSA/FC approach predicts a
tricritical point for the ideal Ising fluid at $H^\ast=0$ and not a critical
end point besides a gas-liquid critical point. This holds for $z_1^\ast \le
5$ and at least to within the relative accuracy $10^{-6}$ of the numerical
calculations.

The change in $z_1^\ast$ (this quantity will be denoted below simply as
$z^\ast$) does not also affect the tendency of the critical temperature
(density) to decrease (increase) monotonically with increasing the external
field strength $H^\ast$. The dependencies $T_{\rm c}^\ast(H^\ast)$ and
$\rho_{\rm c}^\ast(H^\ast)$ are plotted in detail in subsets (a) and (b) of
Fig.~3, respectively. The OZ/SMSA/FC calculations show that in the limit of
weak fields, the functions $T_{\rm c}^\ast(H^\ast)$ and $\rho_{\rm
c}^\ast(H^\ast)$ can be cast in the forms $T_{\rm c}^\ast(H^\ast)=T_{\rm
c}^\ast(0)-c_T(z^\ast)(H^\ast)^{2/5}$ and $\rho_{\rm c}^\ast(H^\ast)=
\rho_{\rm c}^\ast(0)+c_\rho(z^\ast)(H^\ast)^{2/5}$ with $c_T(z^\ast)$ and
$c_\rho(z^\ast)$ being quantities depending only on $z^\ast$. The exponent of
this power law behavior is in accordance with the mean field tricritical
exponent \cite{Schinagl,Fefomel} (where, of course, $c_T$ and $c_\rho$ are
independent of $z^\ast$). With increasing $H^\ast$, the functions $T_{\rm
c}^\ast(H^\ast)$ and $\rho_{\rm c}^\ast(H^\ast)$ begin to tend rapidly
(especially at small and moderate values of $z$) to their infinite field
limits. For larger $z^\ast$, the saturation regime shifts to higher values of
$H^\ast$. At small inverse screening length ($z^\ast< 0.5$), the OZ/SMSA/FC
and SCMF results are practically indistinguishable.

It is worth mentioning that in the Kac limit $z^\ast\to 0$, the
SCMF theory should lead to exact results provided the equation of
state of the reference system is chosen exactly too. Indeed, at
$z^\ast \to 0$ the magnitude $2 (z \sigma)^2/ (z\sigma+1)$ of
magnetic potential vanishes, whereas the screening length $1/z$
tends to infinity. Under such conditions, the magnetic
interactions can be treated as an infinitesimally small
perturbation to the reference potential and the assumptions of the
MF theory become exact. With increasing $z^\ast$, the precision of
the SCMF description goes down. Note that within the standard HSMF
and SCMF theories, the reference system relates exclusively to
nonmagnetic HS or SC repulsion. This is acceptable for long ranged
($z^\ast \lesssim 1$) potentials. When the screening radius is
short enough ($z^\ast \gtrsim 2$), a more appropriate choice of
the reference system (with including a part of the Yukawa
potential) should be performed. On the other hand, the accuracy of
the OZ/SMSA/FC approach when evaluating $T_{\rm c}^\ast$ and
$\rho_{\rm c}^\ast$ is expected to be of order of 10--15\%, in a
wide region of $z^\ast$-values. This is the same accuracy as for
the case $z^\ast=1$, where the direct comparison with the MC data
could be performed (see Fig.~1).

The OZ/SMSA/FC results for the magnetic phase transition (which takes place
only at $H^\ast=0$) are shown in Fig.~4 (a) for the set $z^\ast=0.5$, 1, 2,
3, and 5 of the inverse screening length over a wide temperature and density
region. It can be seen that the dependence of the Curie temperature
$T_\lambda^\ast$ on $\rho^\ast$ shifts considerably to smaller values of
$\rho^\ast$ with increasing $z^\ast$. This dependence is nonlinear contrary
to the HSMF prediction, where  the function $T_\lambda^\ast=8 \pi \rho^\ast$
behaves linearly on $\rho$ and independently of $z$. Within the more accurate
SCMF and OZ/SMSA/FC approaches, the linear dependence of $T_\lambda^\ast$ on
$\rho^\ast$ is recovered only in the particular (Kac) limit $z^\ast \to 0$.
At larger values of $z^\ast$ (namely, at $z^\ast>2$), the deviations from the
limiting behavior become significant. A similar, but considerably weaker
$z^\ast$-dependence of $T_\lambda^\ast$ to that presented in Fig.~4 (a) for
the OZ/SMSA/FC approach is observed within the SCMF theory, due to the
existence of factor $\gamma(T^\ast,z^\ast)$ in the temperature
$T_\lambda^\ast=8 \pi \gamma \rho^\ast$ (see Ref.~\cite{Fefomel}). Note also
that the difference between the SCMF and OZ/SMSA/FC functions
$T_\lambda^\ast(\rho^\ast)$ increases with rising $z^\ast$. However, even for
small values of $z^\ast$, where the $z^\ast$-dependence is not so pronounced,
the SCMF theory leads to worse predictions. This is demonstrated in Fig.~4
(b) for a particular case $z^\ast=1$ by comparison with MC results. Although
here the both theories agree quite well, the SCMF deviations from the MC data
are, nevertheless, slightly larger than those of the OZ/SMSA/FC theory.

\subsection{Results for nonideal models $(0 < R < \infty)$}

\subsubsection{Zero magnetic field}

The phase coexistences of the nonideal Ising fluid with $z_1^\ast=
z_2^\ast=1$ are shown in Fig.~5 for $H^\ast=0$, when the ratio $R$ of
strengths of the magnetic to nonmagnetic interaction is not too small. As can
be seen, all the curves exhibit a tricritical point behavior -- type I of the
thermodynamic phase diagrams -- of the same topology as the case $R=\infty$.
Note that because of the great number of curves, the magnetic phase
transition lines have been omitted in Fig.~5 as well as in Figs.~6--9. They
are presented in detail in Fig.~14 below.

With further decreasing $R$, the shape of the phase diagrams changes in a
characteristic way. This is illustrated in subsets (a) and (b) of Fig.~6.
Beginning from the upper boundary value $R=R_{u}=0.215$, beside the
tricritical point (TCP) a gas-liquid critical point (GLCP) appears in the
paramagnetic phase region (at smaller densities than the tricritical point
density) indicating that the nonmagnetic interaction is strong enough to
condense here the system into the liquid phase. In addition, a triple point
becomes visible. The TCP now corresponds to a liquid-liquid transition
between paramagnetic and ferromagnetic phases. This is different from the
phase behavior of type I in the region $R_{u} < R \le \infty$, where the TCP
relates to the transition between a paramagnetic gas and a ferromagnetic
liquid phase. Such a new topology of the phase behavior below $R_u$ will be
referred to as type II. The appearance of the additional critical point with
decreasing $R$ is explained by an increased weight of nonmagnetic attractions
in the system. The nonmagnetic interaction is sufficiently strong to produce
a gas-liquid transition before the liquid becomes ferromagnetic.

If the nonmagnetic interaction becomes too strong, namely, when the value of
$R$ is below the lower boundary level $R_{l}$, i.e., $R < R_{l}=0.14$,  the
TCP disappears and transforms into a critical end point (CEP) (see subset (b)
of Fig.~6). At the same time, the GLCP remains and further shifts away from
the CEP. Such a topology of the phase diagram in the region $R < R_{l}$ will
be defined as of type III. Note that for extremely small values of $R$ (when
$R \ll R_{l}$), the phase coexistence will behave like that inherent in a
simple nonmagnetic fluid (because then $I \gg J$ and magnetic interactions
can be ignored completely).

These three phase diagram
topologies at $H=0$ have been found earlier for other systems,
such as symmetric binary nonmagnetic mixtures \cite{Wilding,Kahlang,%
Kahl,Paschinger}, the Heisenberg fluid \cite{Schinagl,Weis} or the
Stockmayer fluid \cite{Groh}. It is interesting to remark that the
boundary values $R_{u}=0.215$ and $R_{l}=0.14$ calculated by us
within the OZ/SMSA/FC approach for the Ising fluid correspond to
$\delta_{u}=0.646$ and $\delta_{l}=0.754$. Here $\delta=
(1-R)/(1+R)$ denotes the ratio of interparticle potentials
(outside the hard or soft core) between unlike and like (see
Eq.~(6), $\delta$ is proportional to the ratio
$[I(r)-J(r)]/[I(r)+J(r)]$) particles in the mixture. The latter
values are very close to those ($\delta_{u}=0.65$ and
$\delta_{l}=0.75$) reported in Ref.~\cite{Paschinger} for a
symmetric binary mixture and evaluated within the SCOZA technique
(the HRT yields $\delta_{u}=0.665$ \cite{Pini}). However, the
direct comparison is not possible since we used the soft core
potential instead of the hard sphere repulsion and another value
of the inverse screening length $z \equiv z_1=z_2$.

Varying the parameter $z^\ast$ can lead to a qualitative modification of the
phase diagrams and thus to a shift of the boundaries between different
topologies. This is seen in Fig.~7, where the cases $z_1^\ast =z_2^\ast=0.5$,
$1$, $2$, and $3$ are considered at $R=0.215$ (subset (a)) and $R=0.14$
(subset (b)). From the topologies of these diagrams it can be concluded that
the upper $R_u(z^\ast)$ and lower $R_l(z^\ast)$ boundary values decrease with
increasing $z^\ast \equiv z_1^\ast=z_2^\ast$. A more complicated situation
arises when $z_1^\ast \ne z_2^\ast$ that is presented in Fig.~8. Here, the
quantities $R_u$ and $R_l$ should be treated as depending on both inverse
screening lengths $z_1^\ast$ and $z_2^\ast$. Analyzing the set of curves in
Fig.~8, it can be stated that the behavior of $R_u(z_1^\ast,z_2^\ast)$ and
$R_l(z_1^\ast,z_2^\ast)$ is not monotonic. In particular, the functions
$R_u(z_1^\ast,z_2^\ast)$ and $R_l(z_1^\ast,z_2^\ast)$ increase with rising
$z_2^\ast$ at fixed $z_1^\ast=1$, but they decrease with increasing
$z_1^\ast$ at constant $z_2^\ast=1$.

It is worth mentioning that the above three types of the phase diagrams can
also be observed within MF theory \cite{Schinagl}. The disadvantage of the MF
description is that it produces boundary values $R_u$ and $R_l$ which are
independent of $z_1^\ast$ and $z_2^\ast$. This corresponds, in fact, to the
limiting behavior of $R_u$ and $R_l$ at $z_1^\ast,z_2^\ast \to 0$.

\subsubsection{Nonzero magnetic field}

A set of phase diagrams for different values of the external field
$H$ and relative strength $R$ of internal magnetic to nonmagnetic
interactions are plotted in Fig.~9. Here, we can see that the
change in $H$ modifies considerably the phase coexistence curves
in the nonideal Ising system. At large values of $R$, such
modifications are similar to those of the ideal Ising fluid
(compare, for example, the subset (a) of Fig.~9, $R=5$, with the
subset (b) of Fig.~2, $R=\infty$). Since there is no magnetic
phase transition at finite fields, the tricritical point at
$H^\ast=0$ transforms into a critical point for $H^\ast \ne 0$
(note that one has a phase diagram symmetric in $H\to -H$) and
moves monotonically in the temperature-density plane with
increasing $H$ to the side of lower $T^\ast$ and higher
$\rho^\ast$. This demonstrates that the critical lines meet in a
tricritical point. At intermediate and small values of $R$, the
phase diagram modifications exhibit nonmonotonic features. For
instance, at $R=0.29$ (see subset (c)), the critical temperature,
starting from the value $T_{\rm c}(0)$ at $H=0$, begins first to
go down, reaching a minimum at $H^\ast \sim 2$. Further it
increases up to its limiting value $T_{{\rm c} \infty}=\lim_{H \to
\infty} T_{\rm c}(H)$, where $T_{{\rm c} \infty}$ can be less
(subset (c)) or greater (subset (d)) than $T_{\rm c}(0)$.

More complicated scenarios are observed for parameters
$R_{l}=0.14<R<0.215=R_{u}$ (the region of topologies of type II), when the
gas-liquid critical point exists simultaneously (from the left in
$\rho^\ast$-axis) to the tricritical one. However, only one of these two
critical points at $H=0$ can be connected by a critical line with the
critical point at $H=\infty$. With increasing $H$, one of the two critical
lines has to end. This is only possible in a critical end point. How this
happens depends on the concrete value of $R$ (see subsets (e), (f), and (g)
of Fig.~9). In the infinite field limit $H \to \infty$, the phase diagrams
tend at each $R$ to the gas-liquid binodals of a simple nonmagnetic fluid
with the interparticle potential $\phi(r)=\varphi(r)-I(r)-J(r)$. This is
because then all the spins align exactly along the field vector, so that the
product ${\bf s}_i \cdot {\bf s}_j$ will be equal to 1 (see Eq.~(\ref{eq1}))
for any pair of particles. A similar behavior even for finite fields can be
observed for regions with too low values of $R < R_{l}=0.14$ (see subset
(h)), where the influence of magnetic interactions can be neglected ($J \ll
I$).

The dependencies of the gas-liquid critical temperature ${T_{\rm
c}^\ast}^{(gl)}$, the liquid-liquid critical (wing) temperature ${T_{\rm
c}^\ast}^{(w)}$ and the corresponding densities ${\rho_{\rm c}^\ast}^{(gl)}$
and ${\rho_{\rm c}^\ast}^{(w)}$ of the nonideal Ising fluid on the value
$H^\ast$ of the external field are shown in detail in Figs.~10 and 11,
respectively. They cover the whole region of varying $R$ and include all the
three types, I (subset (a)), II (subsets (b) and (c)), and III (subset (d))
of the phase diagram topology. Note that the both phase transitions exist
only in region II of the global phase diagram, whereas the wing line
disappears for $R<R_l$ and the gas-liquid phase transition line does not
appear for $R>R_u$. For type I (subset (a)), the wing line will correspond to
the gas-liquid critical point (no liquid-liquid phase transitions are present
in $R$-regions corresponding to types I and III). As can be seen, the
monotonic decrease of ${T_{\rm c}^\ast}^{(w)}$ with rising $H^\ast$, observed
at $0.5 \le R \le \infty$, gradually transforms into a nonmonotonic function
${T_{\rm c}^\ast}^{(w)}(H^\ast)$, when the parameter $R$ lies in the interval
$[R_{vl},0.4]$ with $R_{vl}=0.196$ (subset (a) of Fig.~10). The position of
the minimum in ${T_{\rm c}^\ast}^{(w)} (H^\ast)$ shifts from $H^\ast \sim 3$
to 1 with decreasing $R$. At $R \ge R_{vl}$, the wing line can exist for
arbitrary fields $0 \le H^\ast \le \infty$. For $R < R_{vl}$, the wing line
terminates at some finite value $H_{ce}^\ast(R)$ (subset (b) of Fig.~10)
until it disappears at $R < R_{l}=0.14$.

The wing line density ${\rho_{\rm c}^\ast}^{(w)}(H^\ast)$ also
exhibits a nonmonotonic field behavior in the interval $0.6 \le R
< 1.25$ with a maximum at $H^\ast \sim 1$ to 2. Outside of this
interval, it increases ($R \ge 1.25$) or decreases ($R < 0.6$)
monotonically (see subsets (a) and (b) of Fig.~11). On the other
hand, for $0.196=R_{vl} < R < R_{u}=0.215$ the gas-liquid critical
point ends in a critical end point at some finite value
$H_{glce}^\ast(R)$ (subset (b) of Figs.~10 and 11). For $R \le
R_{vl}$, the gas-liquid transition line exists for arbitrary
fields $0 \le H^\ast \le \infty$. The critical temperature
${T_{\rm c}^\ast}^{(gl)}(H^\ast)$ of this transition increases
always monotonically with increasing $H^\ast$ (subsets (c) and (d)
of Fig.~10), whereas the critical density ${\rho_{\rm
c}^\ast}^{(gl)}(H^\ast)$ is always a nonmonotonic function
exhibiting a maximum at $H^\ast \sim 3$ to $H^\ast \sim 30$
depending on $R$ (see subsets (c) and (d) of Fig.~11).

\subsubsection{Van Laar point in the global phase diagram}

From the phase diagram analysis presented it follows that in the
region of topologies of type II ($R_{l}<R<R_{u})$, the gas-liquid
and liquid-liquid transitions can coexist simultaneously not only
at $H=0$ but also for $H \ne 0$. On the other hand, it has been
realized that for sufficiently strong fields including the limit
$H \to \infty$, we have a simple nonmagnetic-like phase behavior
with the presence of only one gas-liquid transition (see Fig.~9).
Thus, region II in the global phase diagram has to split into two
subregions in dependence whether the gas-liquid (type IIa, $R_{vl}
< R < R_u$) or the liquid-liquid (type IIb, $R_l < R < R_{vl}$) critical
line terminates in a critical end point at some finite value of
$H^\ast$. The boundary in the global phase diagram between these
two regions defines a van Laar-like point \cite{meijer88}.

Our calculations have shown that this special point is identified
at $R_{vl}=0.196$ with $H_{tr}^\ast \approx 2.2$ for
$\rho_{tr}^\ast \approx 0.55$ and $T_{tr}^\ast \approx 21.3$. As
can be seen in Fig.~12, there exists a some finite critical value
$H_{tr}$, where both the gas-liquid and liquid-liquid critical
lines merge in one point in the temperature-density plane
projection. It is an asymmetric tricritical point and the change
of the critical exponent $\beta$ from the mean field value ($\beta
=1/2$) to its tricritical value ($\beta =1/4$) is seen in Fig.~12
(subset (b)). The van Laar point has been found in symmetric
mixtures \cite{konynenburg} but so far not in Ising liquids. In
van der Waals theory the value of the van Laar point is
$R_{vl}^{\rm MF}=0.279$ in agreement with the value $\delta=0.564$
given in \cite{konynenburg} (note that $\Lambda=1-\delta=0.436$
there). The value $R_{vl}=0.196$ is comparable to that found (at
$z^\ast=1.8$) in \cite{Pini} (their $\delta_{vl}=0.67$ corresponds
to $R_{vl}=0.197$). In order to collect our results for the
thermodynamic phase diagrams, the global phase diagram in
$R$-space is presented in Fig.~13.

\subsubsection{Magnetic critical line}

Let us consider, finally, the OZ/SMSA/FC result on the para-ferro coexistence
in the nonideal Ising fluid at $H=0$. It is shown in Fig.~14 for various
values of $R$ in a wide region of the temperature-density plane. In the case
of the ideal fluid ($R=\infty$), a strong dependence of the Curie temperature
$T_\lambda^\ast$ on the screening length of the magnetic interaction (see
subset (a) of Fig.~4) has been found. As can be observed in Fig.~14, the
function $T_\lambda (\rho)$ exhibits also an $R$-dependence, especially at
low densities. This is in a contrast to predictions of the HSMF and SCMF
theories which lead to values of $T_\lambda^\ast(\rho^\ast)$ independent of
$R$ (see dashed curves in Fig.~14). However at larger densities ($\rho^\ast
>1$) all the OZ/SMSA/FC functions begin to converge to the same curve, which
is almost a straight line, and the $R$-dependence vanishes. This line does
not coincide with the HSMF and SCMF results.

\section{Conclusions}

We have formulated a generalization of the integral equation formalism for
symmetric binary mixtures in order to study phase coexistence properties of
Ising spin fluids in the presence of an external magnetic field $H$. Mapping
the spin system onto the binary mixture shows that the calculations for the
soft-core Ising fluid at a certain magnetic field reduce in the mixture
picture to the calculations at a certain value of the chemical potential
difference of the constituents of the mixture. This introduces a field
constraint to the OZ equations and modifies the MSA closure to the SMSA
ansatz. It has been demonstrated that the resulting OZ/SMSA/FC approach is
able to describe adequately the phase behavior of such models. Depending on
the ratio $R$ of the strengths of the magnetic and nonmagnetic interactions
inherent in the spin fluid system, four types of thermodynamic phase diagrams
have been identified.

As has been established, the OZ/SMSA/FC approach provides us with
more accurate predictions in comparison to those of the MF theory
and corroborates the MF global phase diagram, changing of course
the boundary values of $R$ which separates the different
topologies. It is expected that some other more complicated
schemes, such as the SCOZA, for example, should lead to a higher
precision of the calculations. However, they are not yet developed
in their present formulations to be directly applied to magnetic
fluids with soft-core repulsion potentials. Due to their high
level of sophistication, they meet considerable computational
difficulties in actual implementations. On the other hand, the
approach proposed here can be used for systems with arbitrary
potentials at relatively low computational costs.

The OZ/SMSA/FC scheme can also be extended to magnetic systems where spins
accept more than two discrete values (they map onto a nonmagnetic
multi-component mixture). The present SMSA can be applied to a soft XY and
Heisenberg fluids (replacing the discrete external field constraint (Eqs.~(9)
and (18)) by its continuous spin counterpart, such as the Lovett equation
\cite{Lovett}, for instance). These questions as well as the problem of
improving the thermodynamic self-consistency of the integral equation
approach is left for future considerations.

\begin{center}
{\small \bf ACKNOWLEDGMENT}
\end{center}

Part of this work was supported by the Fonds zur F\"orderung der
wissenschaftlichen Forschung under Project No.~P15247. I.O. and I.M. thank
the Fundamental Researches State Fund of the Ministry of Education and
Science of Ukraine for support under Project No. 02.07/00303.

\newpage

\begin{center}
FIGURE CAPTIONS
\end{center}

\vspace{6pt}

FIG.~1. The liquid-gas coexistence densities $\rho^\ast$ as a function of
temperature $T^\ast$ obtained within the OZ/SMSA/FC integral equation
approach (bold curves) for the soft-core ideal ($R=\infty$) Ising fluid with
$z_1^\ast=1$ and different values of the external field, $H^\ast= 0$, 0.1,
0.5, 1, 5, and $\infty$,  (subsets (a), (b), (c), (d), (e), and (f),
respectively). The results of the usual and the adjustable versions
\cite{Fefomel} of the SCMF theory are plotted correspondingly by thin and
dashed curves. The Gibbs ensemble MC simulation data \cite{Fefomel} are shown
as circles. The para-ferro magnetic phase transition (at $H^\ast=0$, subset
(a)) is shown by the long dashed curve.

\vspace{6pt}

FIG.~2. The liquid-gas coexistence densities $\rho^\ast$ as a function of
temperature $T^\ast$ obtained within the OZ/SMSA/FC integral equation
approach for the soft-core ideal Ising fluid at different values,
$z_1^\ast=0.5$, 1, 2, and 3, of the inverse screening length (subsets (a),
(b), (c), and (d), respectively) as well as different values of the external
field, namely, top to bottom and alternating solid and dashed curves,
$H^\ast= 0$, 0.01, 0.1, 0.3, 0.5, 1, 2, and 5 (within each subset). At
$z_1^\ast=3$, an additional curve corresponding to $H^\ast= 9$ is included in
subset (d). Note that the curves corresponding to $H^\ast = \infty$ are not
shown, because they practically coincide with those related to $H^\ast=5$ at
$z_1^\ast=0.5$, 1, and 2 (subsets (a), (b), and (c)) or to $H^\ast=9$ at
$z_1^\ast=3$ (subsets (d)). The para-ferro magnetic phase transition (at
$H^\ast=0$) is plotted in the subsets by the long dashed line.

\vspace{6pt}

FIG.~3. The critical temperature $T_{\rm c}^\ast$ (subset (a)) and critical
density $\rho_{\rm c}^\ast$ (subset (b)) as functions of the external field
$H^\ast$, evaluated within the OZ/SMSA/FC approach for the ideal Ising fluid
at various values of the inverse screening length $z^\ast$. At $z^\ast \to
0$, the result corresponds to the SCMF theory.

\vspace{6pt}

FIG.~4. (a) The para-ferro magnetic transition temperature $T_\lambda^\ast$
obtained at $H^\ast=0$ within the OZ/SMSA/FC integral equation approach for
the soft-core ideal Ising fluid for different values of inverse screening
length, from right to left, $z^\ast=$0.5, 1, 2, 3, and 5. The result of the
standard HSMF theory is plotted as the dashed straight line and relates to
the case $z^\ast \to 0$. (b) The results of the SCMF and OZ/SMSA/FC theories
for the case $z^\ast=1$ are shown as dashed and solid curves, respectively,
in comparison with canonical MC simulation data (circles) taken from Ref.
\cite{Fefomel}.

\vspace{6pt}

FIG.~5. The gas-liquid coexistence densities $\rho^\ast$ as a function of
temperature $T^\ast$ obtained within the OZ/SMSA/FC approach for the
soft-core nonideal Ising fluid with $z_1^\ast=z_2^\ast=1$ at $H^\ast=0$ for
larger values of $R$.

\vspace{6pt}

FIG.~6. The coexistence densities $\rho^\ast$ as a function of temperature
$T^\ast$ obtained within the OZ/SMSA/FC approach for the soft-core nonideal
Ising fluid with $z_1^\ast=z_2^\ast=1$ at $H=0$ for moderate (subset (a)) and
small (subset (b)) values of $R$.

\vspace{6pt}

FIG.~7. The coexistence densities $\rho^\ast$ as a function of temperature
$T^\ast$ obtained within the OZ/SMSA/FC approach for the nonideal Ising fluid
at $R=0.215$ (subset (a)) and $R=0.14$ (subset (b)) corresponding to
different values, $z_1^\ast=z_2^\ast=0.5$, 1, 2, and 3, of the inverse
screening radii.

\vspace{6pt}

FIG.~8. The same as in Fig.~7 but for $z_1 \ne z_2$ at $R=0.215$ (subsets (a)
and (b)) and $R=0.14$ (subsets (c) and (d)).

\vspace{6pt}

FIG.~9. The complete thermodynamic phase diagrams of the nonideal Ising fluid
with $z_1^\ast=z_2^\ast=1$ evaluated using the OZ/SMSA/FC approach and
projected onto the ($T^\ast, \rho^\ast$) plane at some typical values of
$R=5$, 1, 0.29, 0.215, 0.2, 0.19, 0.16, and 0.12 (subsets (a), (b), (c), (d),
(e), (f), (g), and (h), respectively). The families of the diagrams in each
of the subsets correspond to different values of the external field,
$H^\ast=0$, 0.01, 0.1, 0.5, 1, 2, 3, 5, 9, and $\infty$ (the value
$H^\ast=0.3$ for $R=5$ and 1, as well as $H^\ast=20$ for $R=0.12$ are
included additionally, whereas $H^\ast=9$ is excluded for $R=5$ and 1).

\vspace{6pt}

FIG.~10. The critical temperature of the wing lines ${T_{\rm c}^\ast}^{(w)}$,
subsets (a) and (b), and the gas-liquid critical lines ${T_{\rm
c}^\ast}^{(gl)}$, subsets (c) and (d), as a function of the external magnetic
field $H^\ast$ for the nonideal Ising fluid at different values of parameter
$R$.

\vspace{6pt}

FIG.~11. The critical density of the wing lines ${\rho_{\rm c}^\ast}^{(w)}$,
subsets (a) and (b), and the gas-liquid critical lines ${\rho_{\rm
c}^\ast}^{(gl)}$, subsets (c) and (d), as a function of the external magnetic
field $H^\ast$ for the nonideal Ising fluid at different values of $R$.

\vspace{6pt}

FIG.~12. The thermodynamic phase diagrams obtained within the OZ/SMSA/FC
approach for the soft-core nonideal Ising fluid with $z_1^\ast=z_2^\ast=1$ at
$R=R_{vl}=0.196$. The phase coexistence curves projected onto the ($T^\ast,
\rho^\ast$) plane are shown in subset (a) for different external field
values, $H^\ast=0$, 0.01, 0.1, 0.5, 1, 2, 3, 5, 9, and $\infty$. A more
detailed phase behavior near the asymmetric tricritical point is presented in
subset (b) for (bottom to top) $H^\ast=1$, 1.2, 1.4, 1.6, 1.8, 2, 2.2, and
2.6. Here, the gas-liquid (on the left) and wing line (on the right) critical
points (circles) are connected by thin curves. The critical temperature
${T^\ast_{\rm c}}^{(gl,w)}$ and critical density ${\rho^\ast_{\rm
c}}^{(gl,w)}$ of the gas-liquid and liquid-liquid phase transitions are
plotted as functions of $H^\ast$ in subset (c) and (d), respectively. They
meet in the tricritical point (dots).

\vspace{6pt}

FIG.~13. Global phase diagram of an Ising fluid. Type I contains
at $H=0$ a tricritical point, where the magnetic transition line
and two wing lines, existing for arbitrary $\pm H$, meet. Type IIa
contains a tricritical point (at $H=0$) and the gas-liquid
critical lines (for $\pm H$), ending in a critical end point at
finite magnetic field, while the wing lines exists for arbitrary
magnetic field. Type IIb contains a tricritical point (at $H=0$)
and the gas-liquid critical lines (for $\pm H$) extending to
infinite magnetic field, whereas the wing lines end in a critical
end point at finite magnetic field. Type III contains only
gas-liquid critical lines extending to infinite values of magnetic
field, and the magnetic transition line ends in a critical end
point at $H=0$.

\vspace{6pt}

FIG.~14. The para-ferro magnetic coexistence curves evaluated at $H=0$ using
the OZ/SMSA/FC theory for the soft-core nonideal Ising fluid with
$z_1^\ast=z_2^\ast=1$ at different values of the system parameter, namely,
top to bottom, $R=0.1$, 0.12, 0.14, 0.17, 0.215, 0.29, 0.4, and $\infty$. The
results of the HSMF and SCMF approaches are plotted as the short- and
long-dashed curves, respectively.

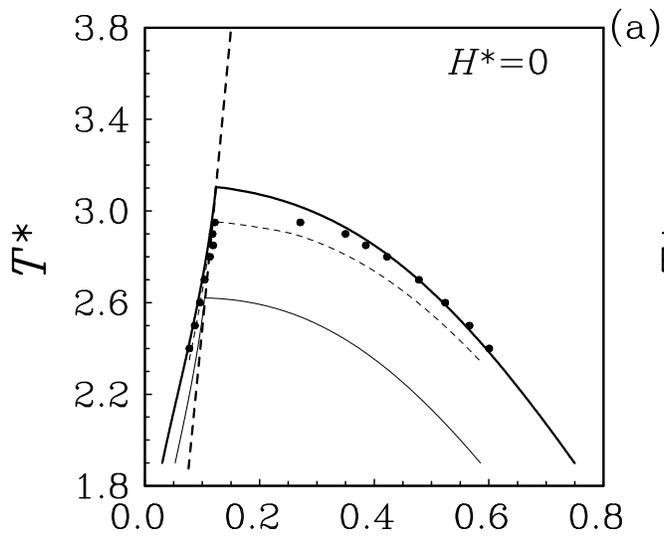
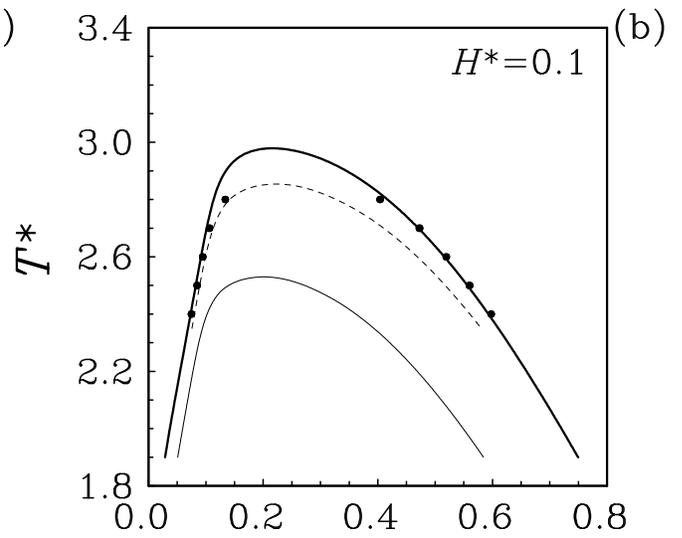
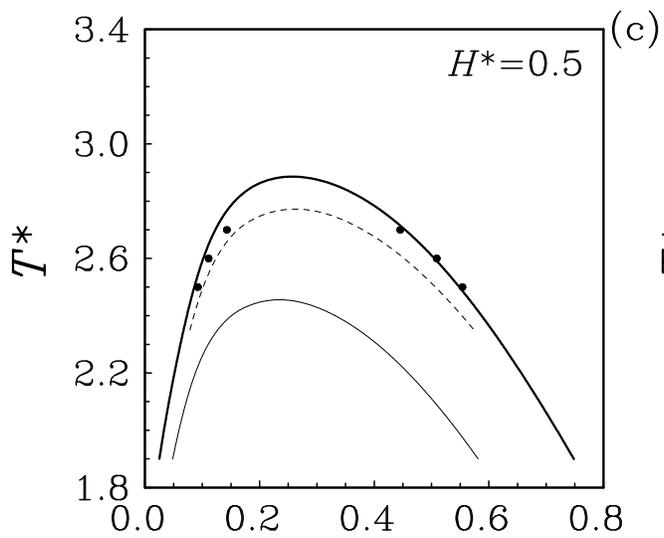
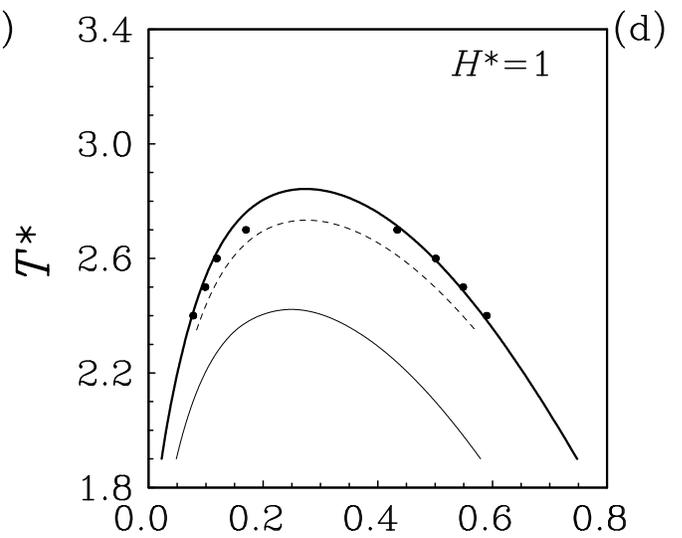
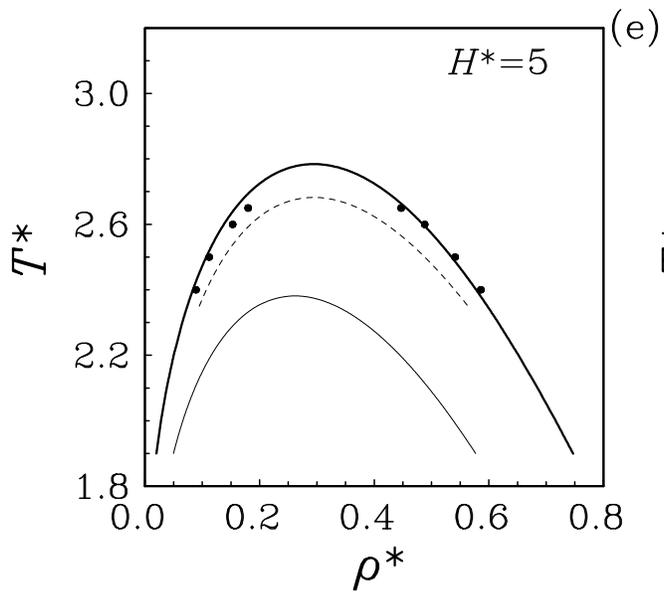
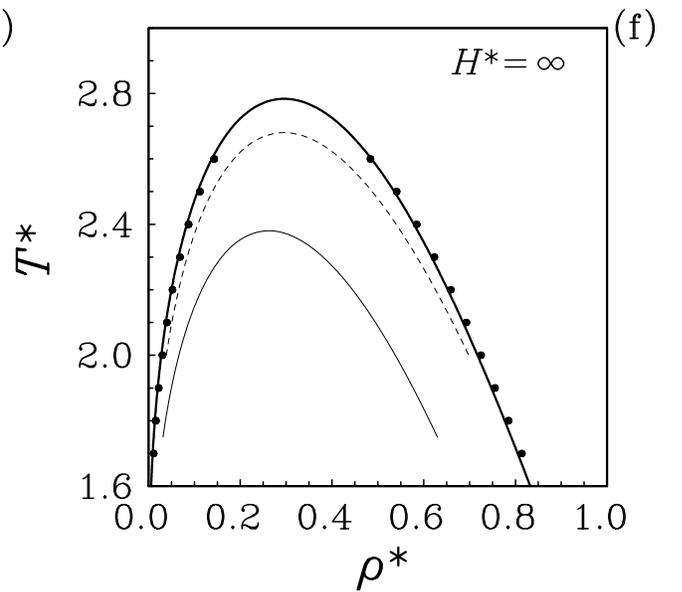
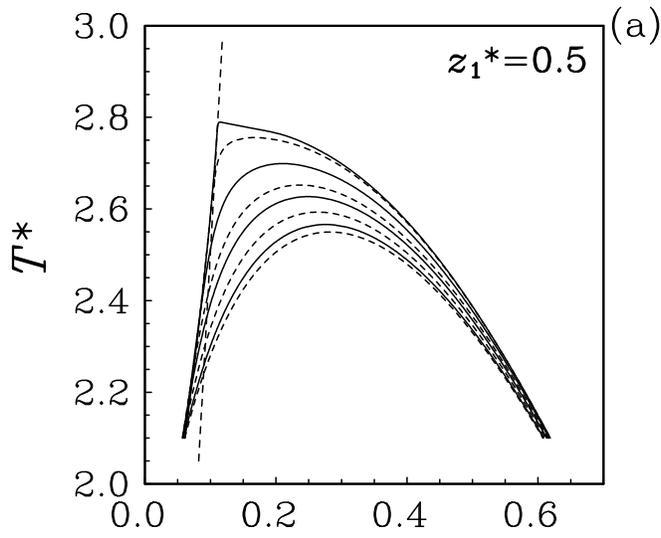
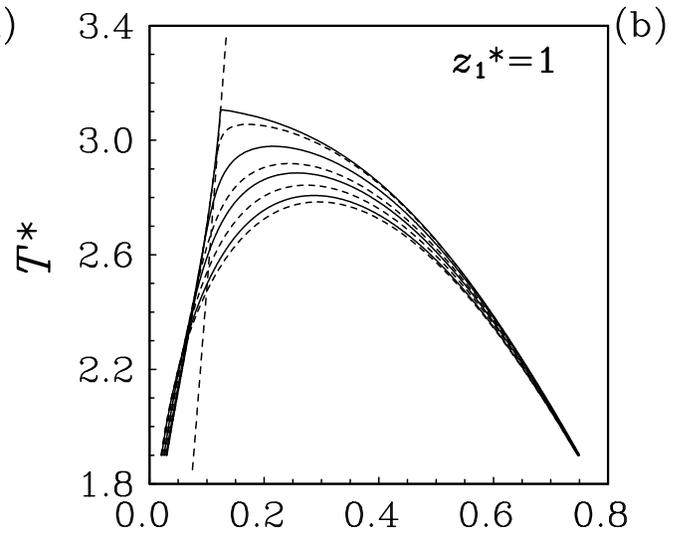
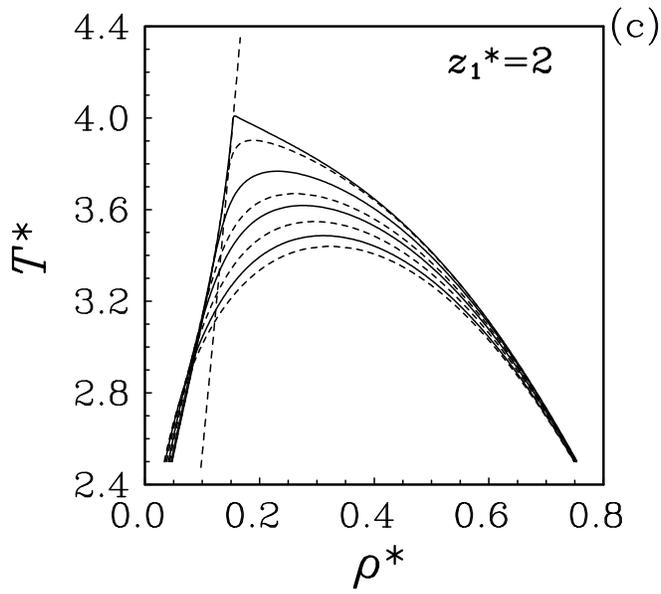
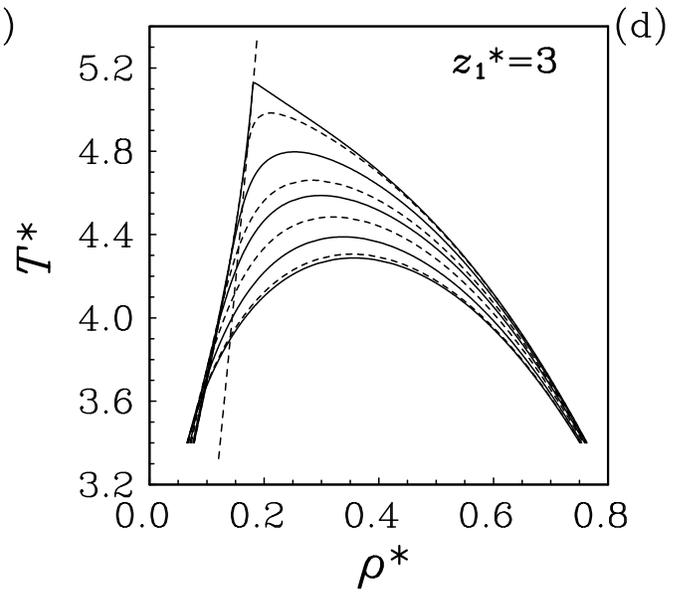
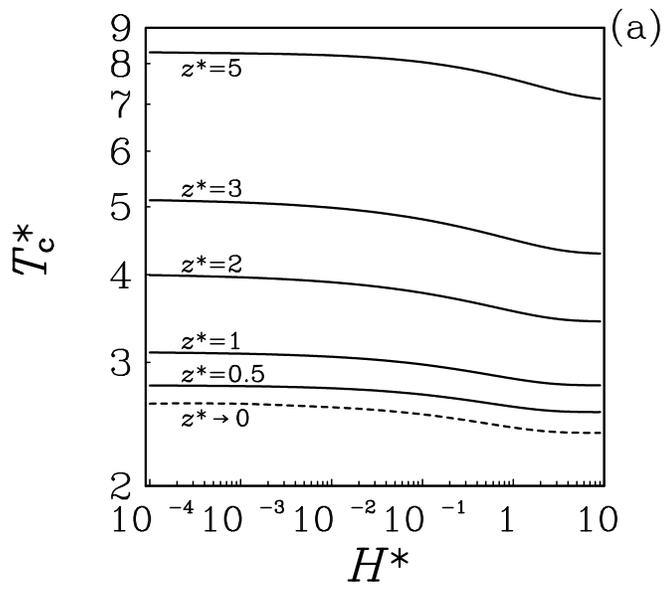
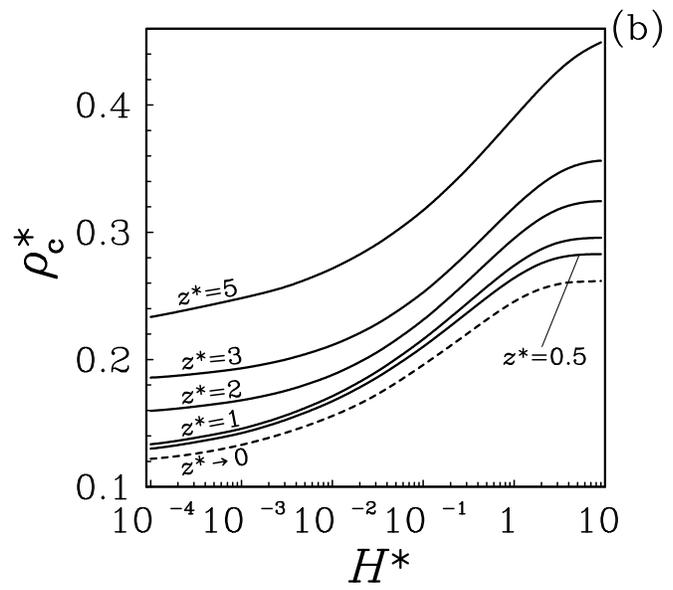
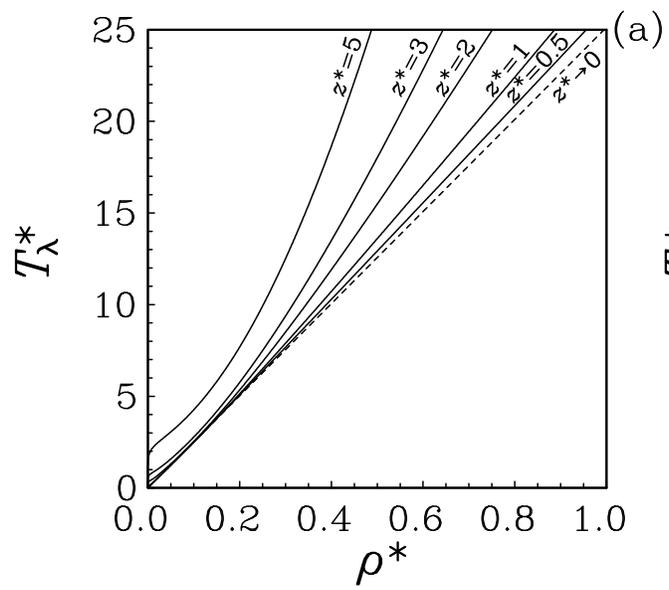
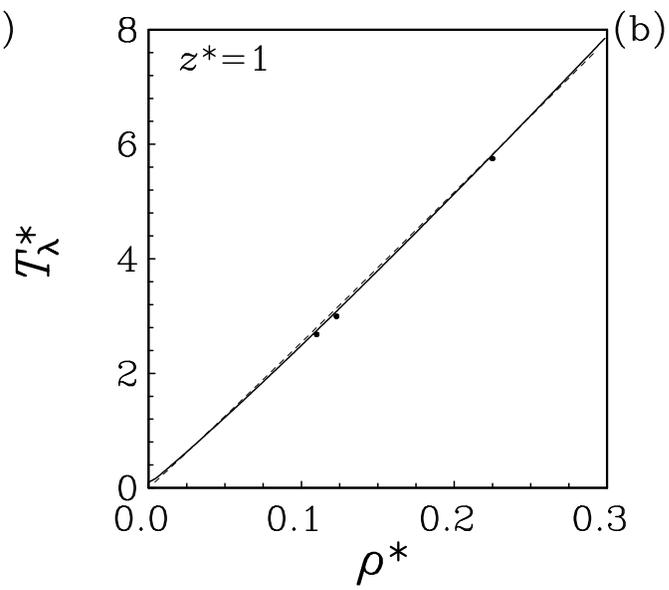
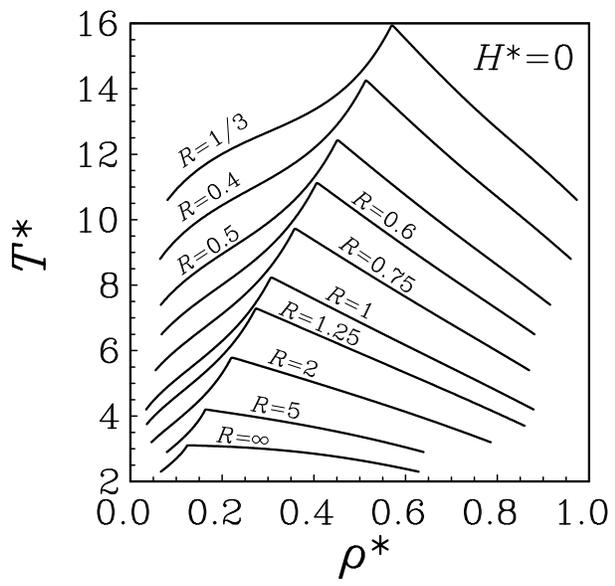
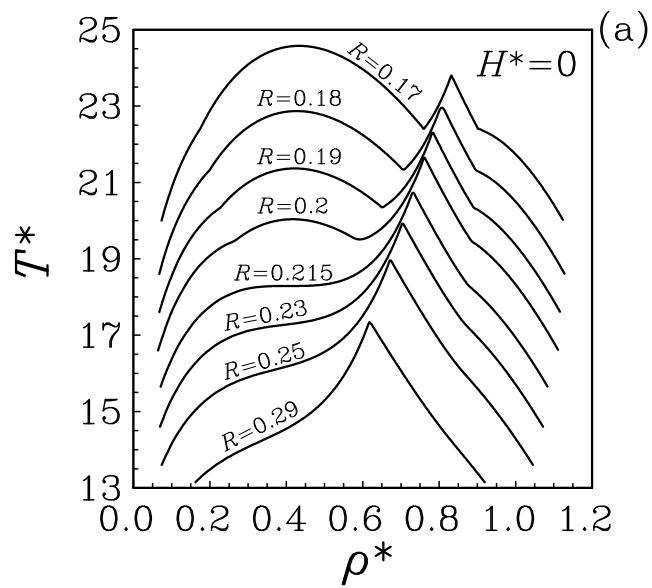
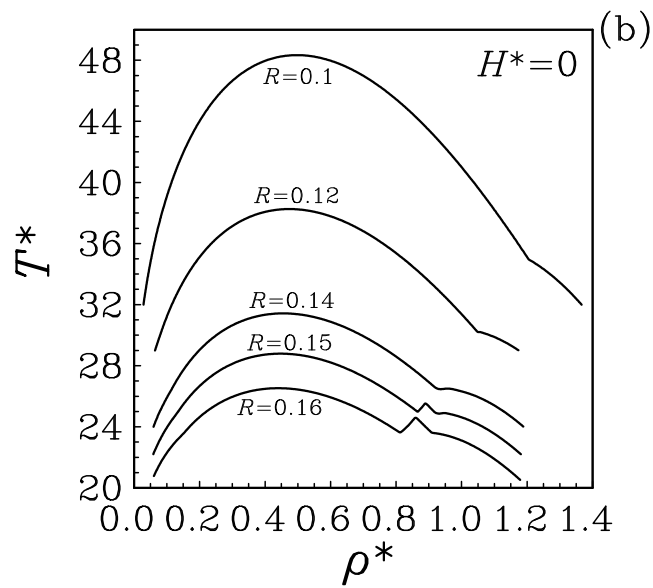
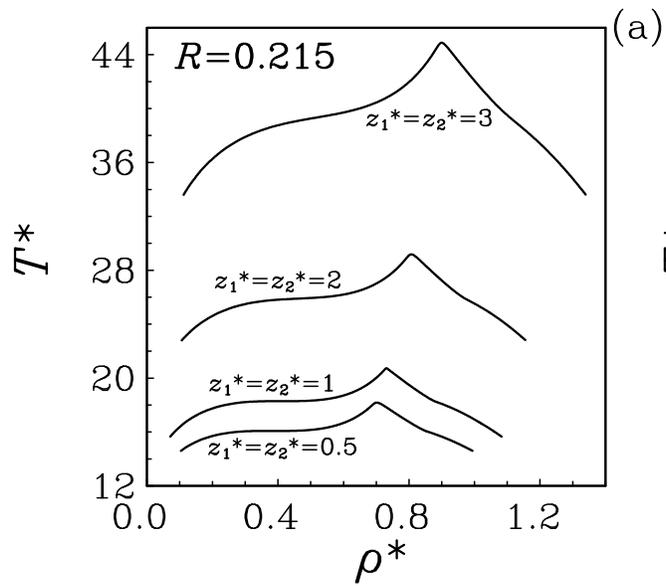
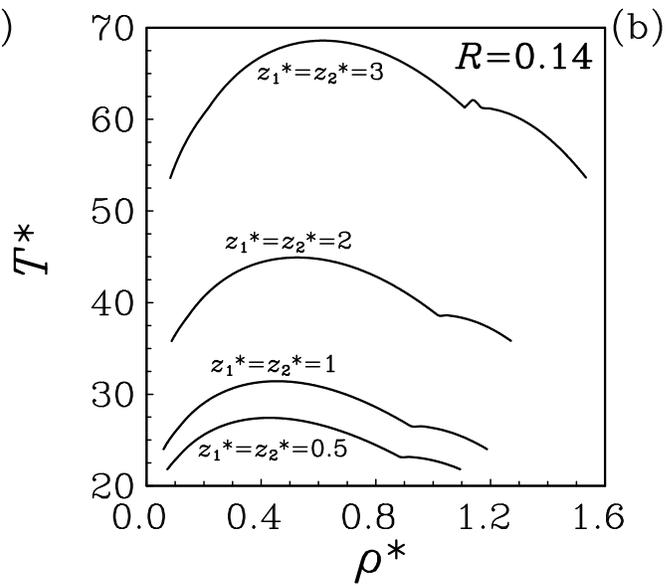
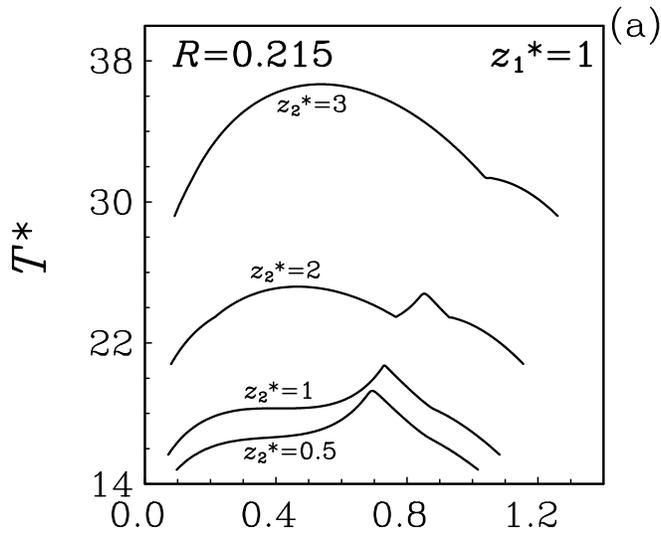
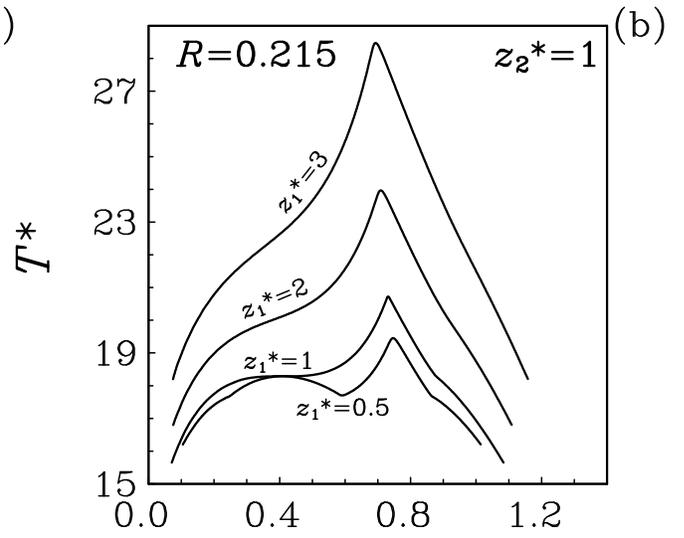
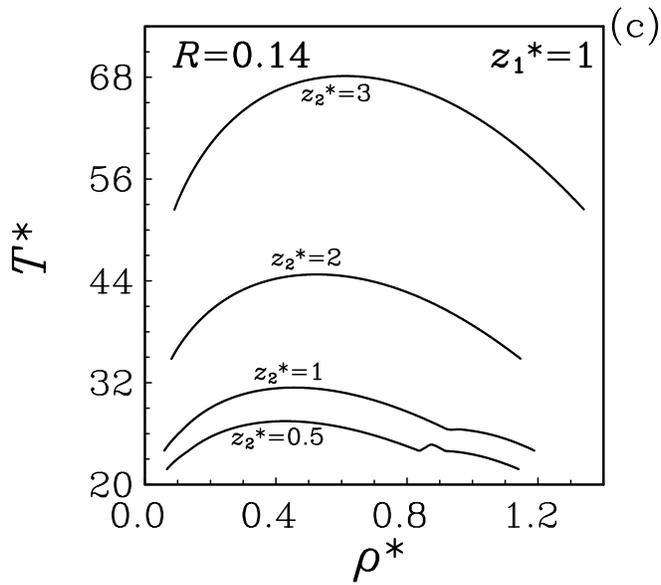
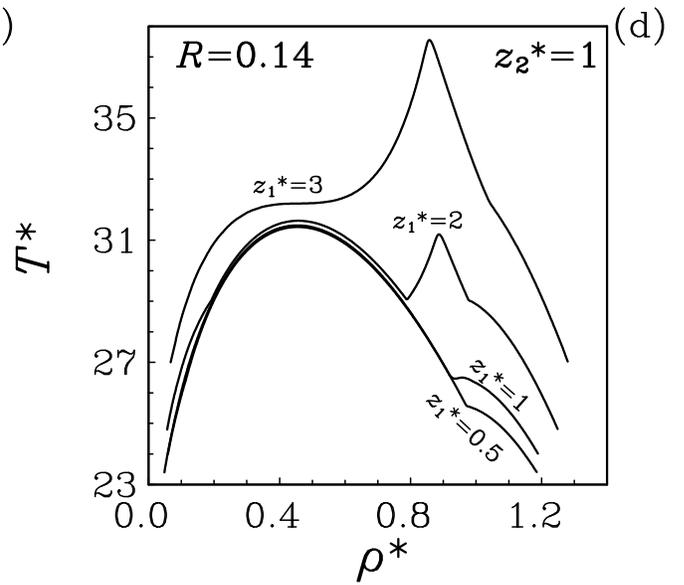
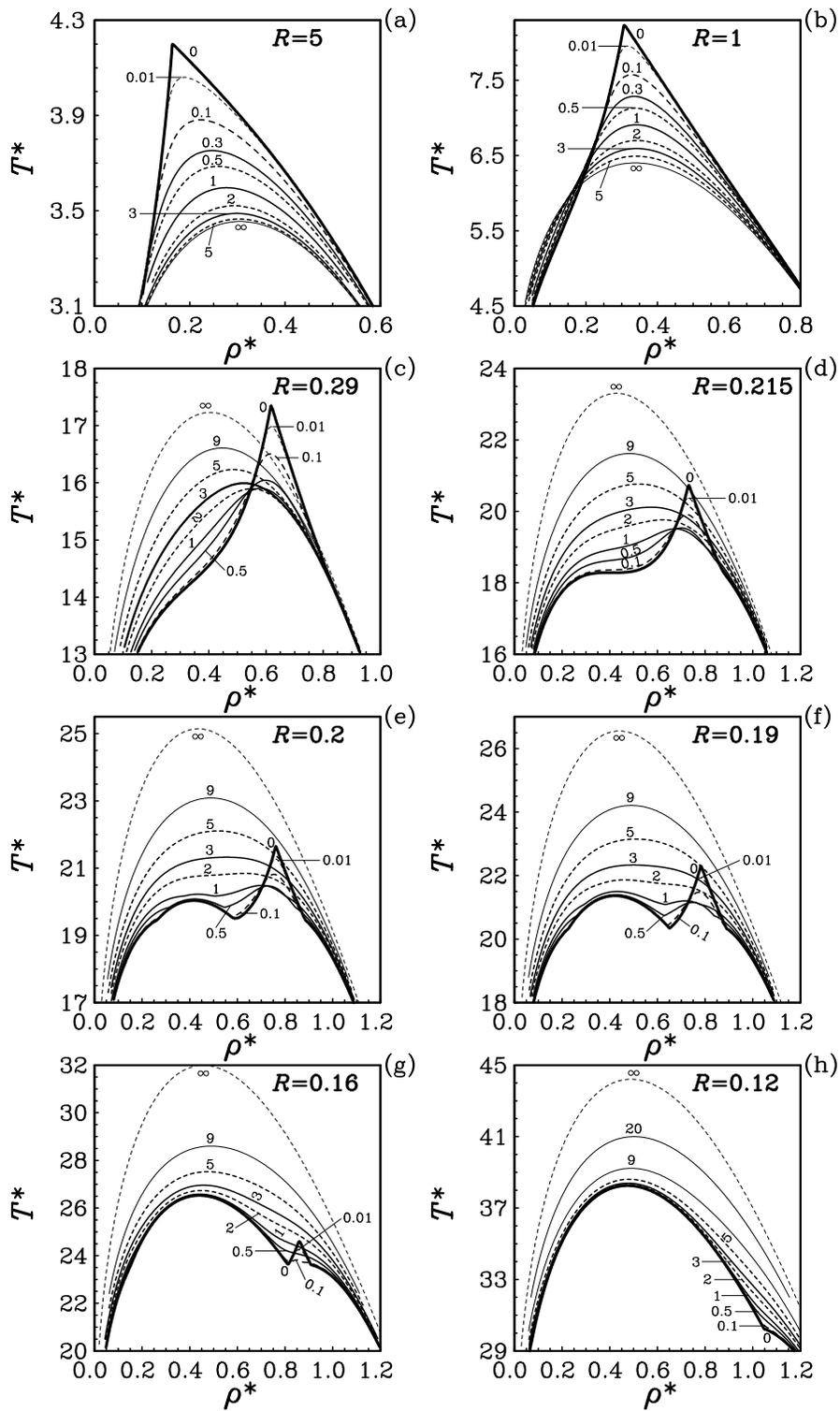
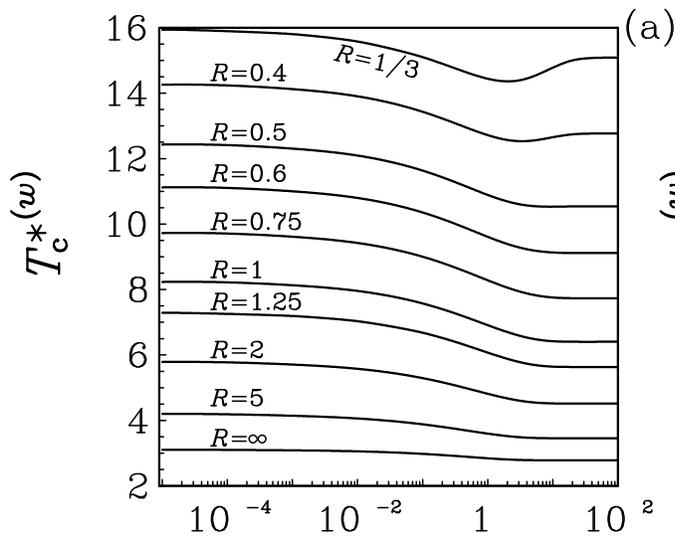
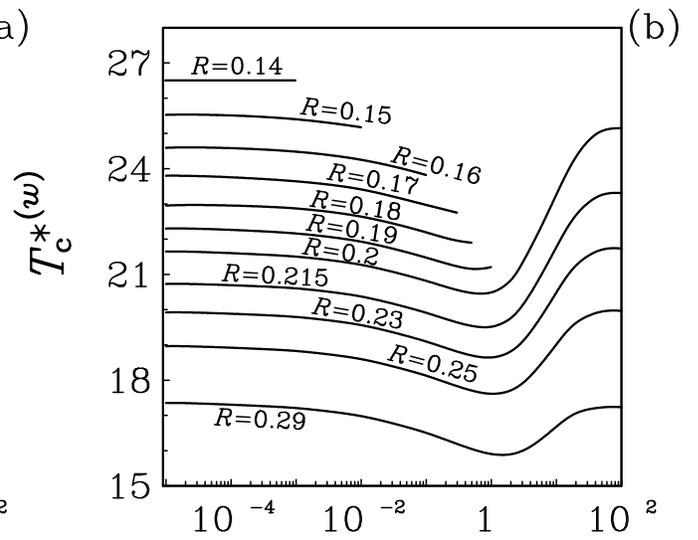
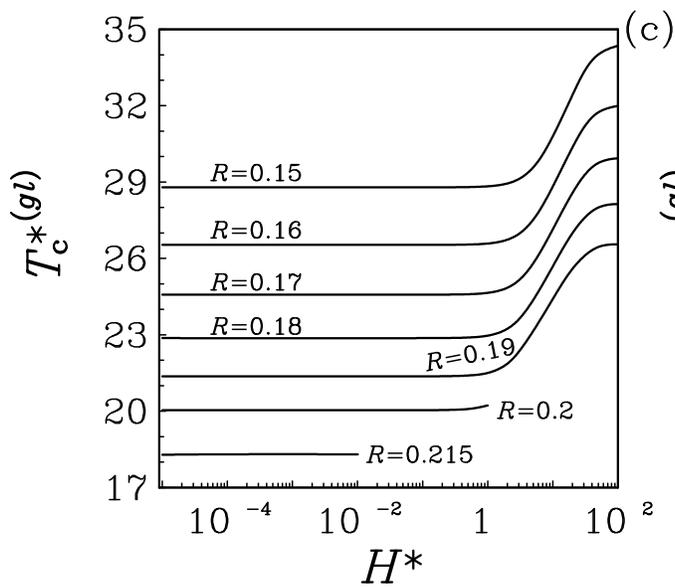
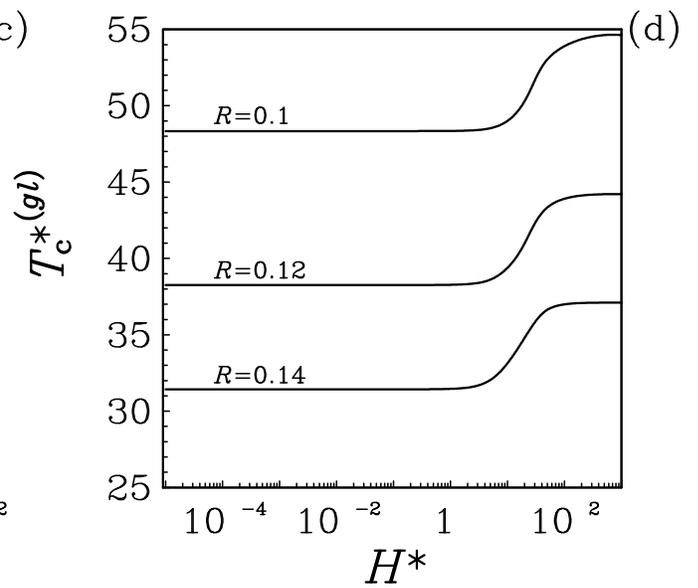
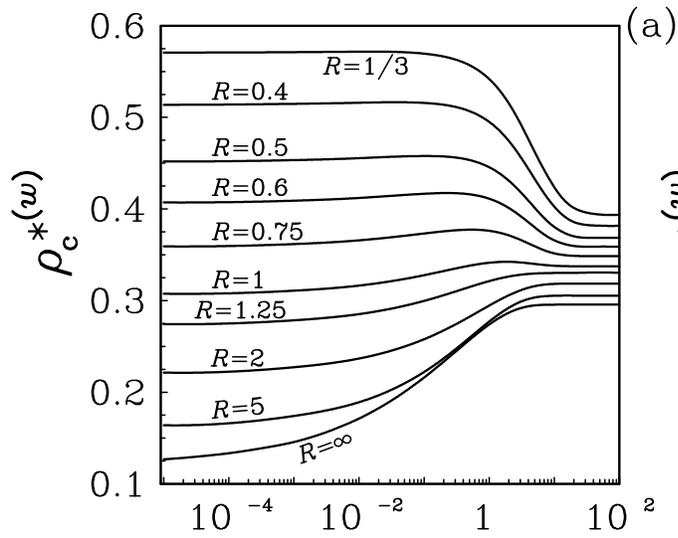
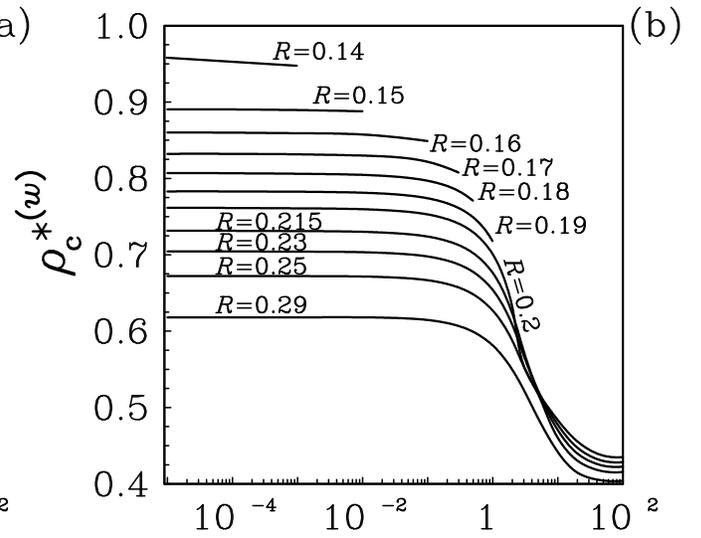
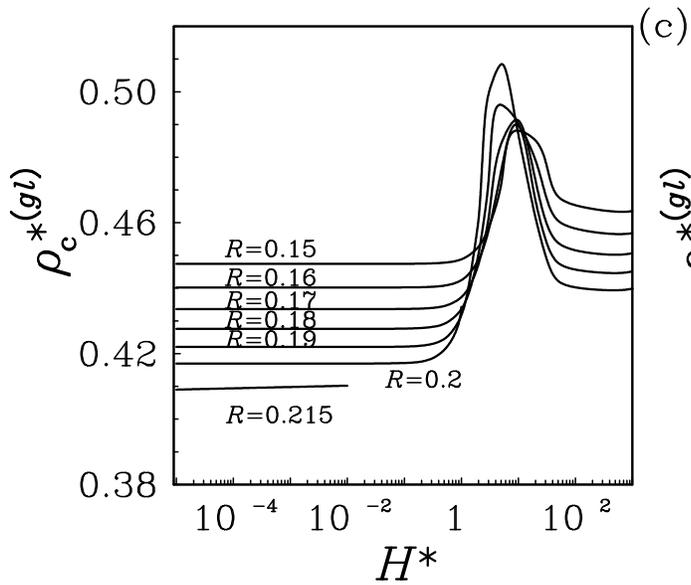
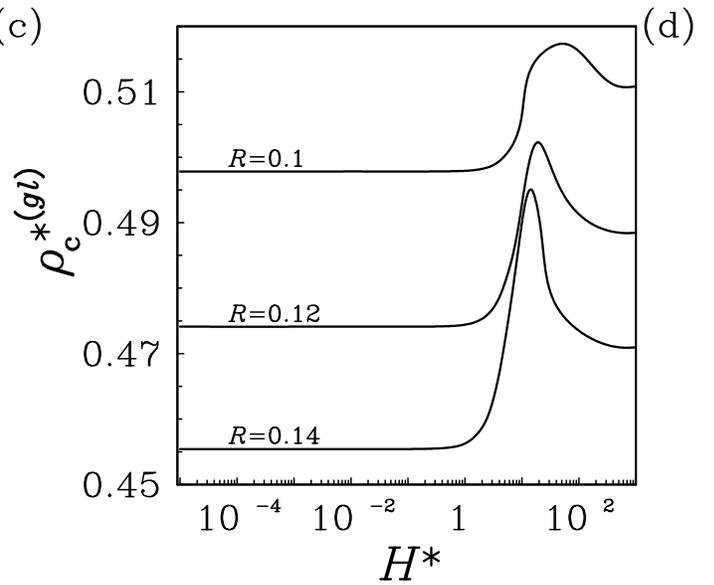
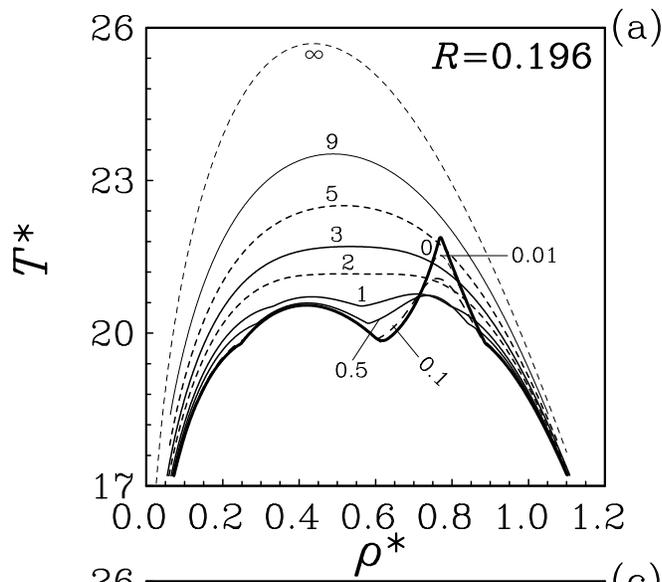
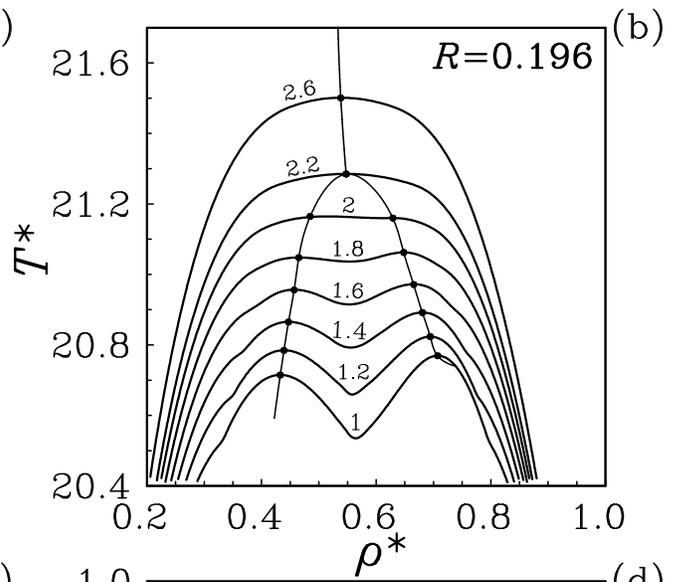
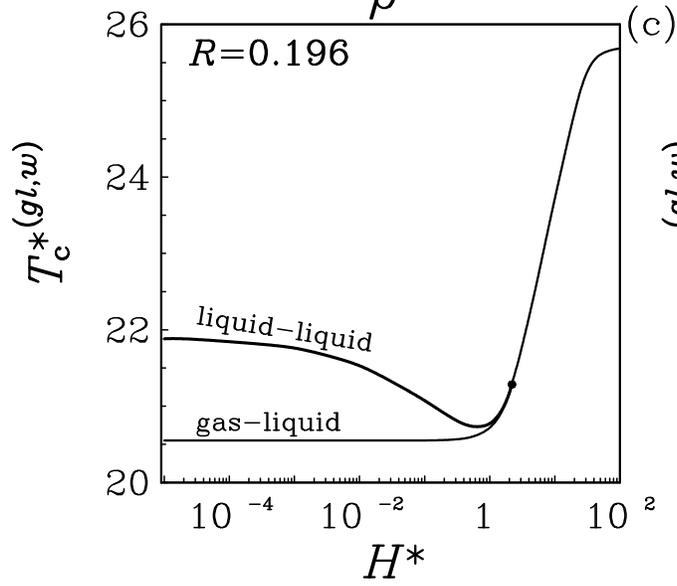
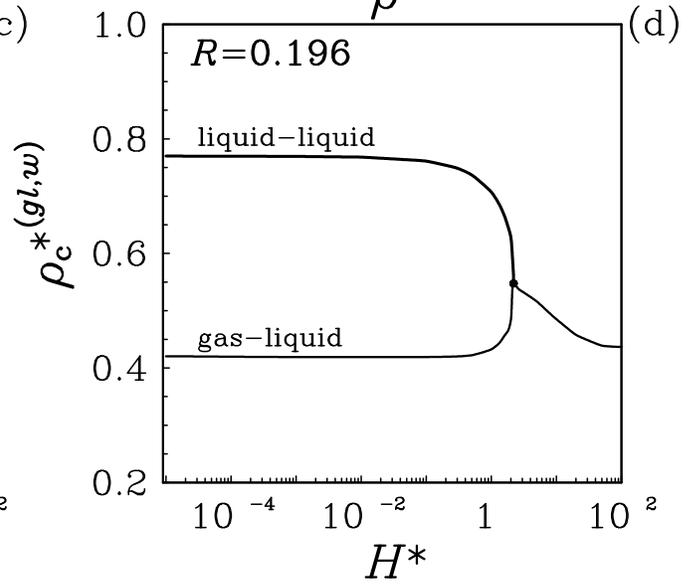
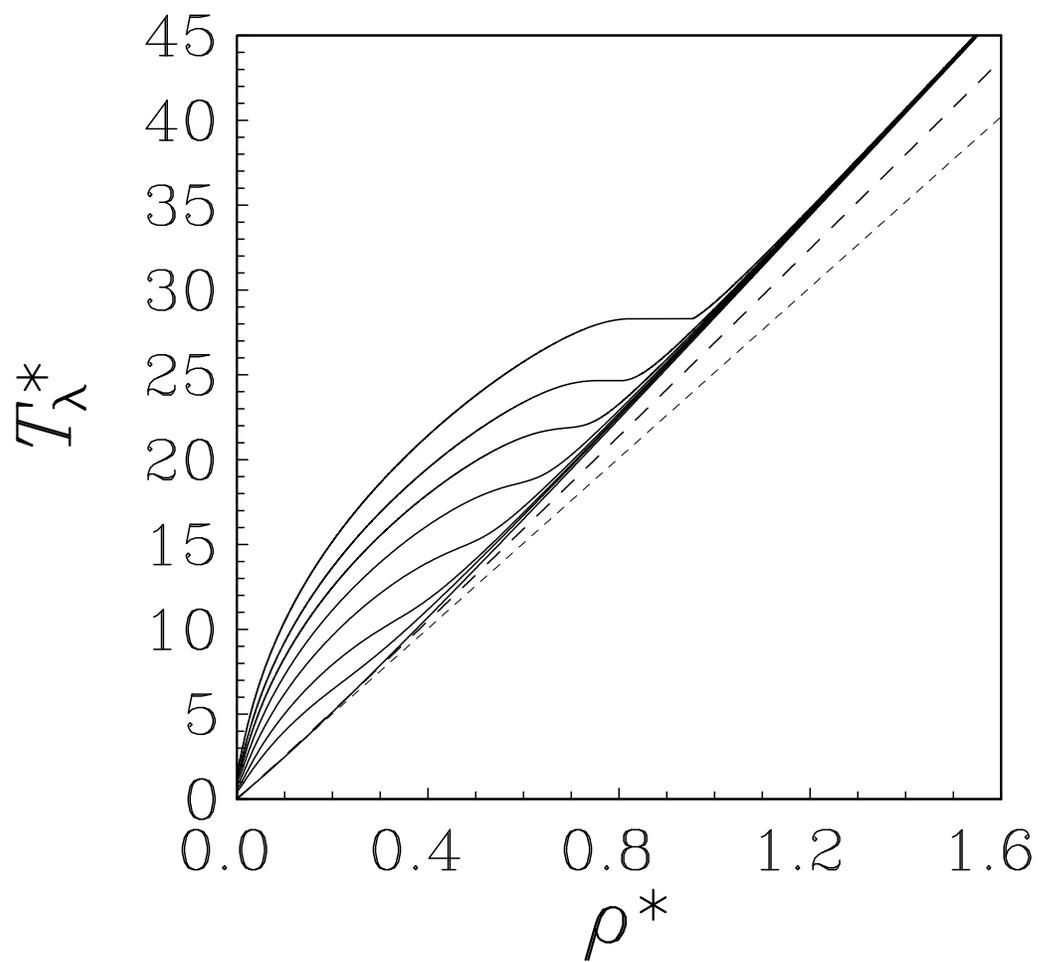
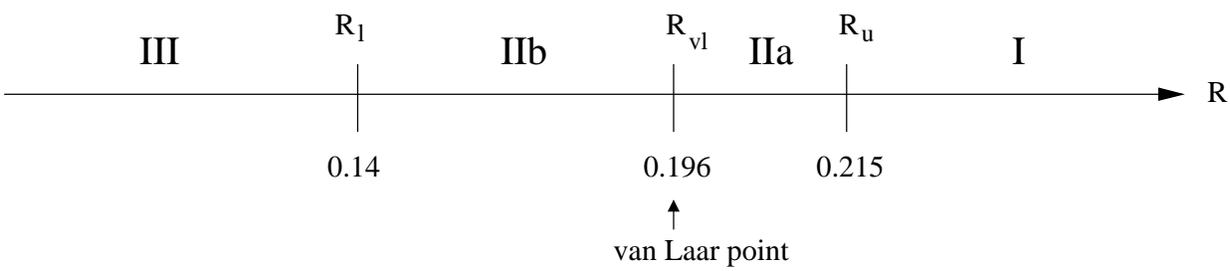
\end{document}